\documentclass[aps,prd,nofootinbib,superscriptaddress,twoside,twocolumn,floatfix,a4paper,reprint,preprintnumbers,showkeys]{revtex4-2}
\usepackage[html]{xcolor}

\usepackage{epsfig,mathtools,bm,color,xcolor,graphicx,braket,adjustbox,esint,upgreek,dcolumn,psfrag}
\usepackage{amsmath,amssymb,amsfonts}
\usepackage[utf8]{inputenc}
\usepackage[english]{babel}
\usepackage{slashed}
\usepackage{orcidlink}
\usepackage{multirow} 

\usepackage{soul,cancel}

\allowdisplaybreaks 
\usepackage{hyperref} 
\hypersetup{
colorlinks,
linkcolor={blue},
citecolor={green!70!black},
urlcolor={blue!80!black},
pdftitle={},
}
\usepackage{cleveref}

\newcommand{\0}{^{\rm{ph}}}
\newcommand{\oo}[1]{(#1^{\rm ph})^2}

\newcommand{\mbar}{\bar{m}_D}

\newcommand{\be}{\begin{equation}}
\newcommand{\ee}{\end{equation}}
\newcommand{\bea}{\begin{eqnarray}}
\newcommand{\eea}{\end{eqnarray}}
\newcommand{\beas}{\begin{eqnarray*}}
\newcommand{\eeas}{\end{eqnarray*}}

\newcommand{\vep}{{\bm p}} 

\newcommand{\veq}{{\bm q}}

\newcommand{\ver}{{\bm r}}

\newcommand{\veep}{{\bm \epsilon}}

\newcommand{\Tcc}{T_{cc}^+}
\newcommand{\Tbb}{T_{bb}^-}
\newcommand{\Tbc}{T_{bc}}
\newcommand{\pcot}{p\cot\delta_0}
\newcommand{\muD}{m_r}
\newcommand{\Eth}{E_{\rm th}}
\newcommand{\Ep}{E^{\rm p}}
\newcommand{\pp}{p^{\rm p}}
\newcommand{\Elhc}{E_{\rm lhc}}
\newcommand{\plhc}{p_{\rm lhc}}
\newcommand{\vp}{{\bm p}}
\newcommand{\vk}{{\bm q}}

\newcommand{\dk}{\frac{d^3q}{(2\pi)^3}}

\def\vec#1{\boldsymbol{#1}}

\graphicspath{{Figs.dir/}}

\newcommand{\ijs}{\affiliation{Jo{\v z}ef Stefan Institute, Jamova 39, 1000, Ljubljana, Slovenia}}

\newcommand{\lis}{\affiliation{CeFEMA, Center of Physics and Engineering of Advanced Materials, Instituto Superior T{\'e}cnico, Av. Rovisco Pais, 1 1049-001 Lisboa, Portugal}}

\newcommand{\reg}{\affiliation{Institut f{\" u}r Theoretische Physik, Universit{\" a}t Regensburg, 93040 Regensburg, Germany}}

\newcommand{\ulj}{\affiliation{Faculty of Mathematics and Physics, University of Ljubljana, 1000 Ljubljana, Slovenia}}

\newcommand{\imsc}{\affiliation{The Institute of Mathematical Sciences, a CI of Homi Bhabha National Institute, Chennai, 600113, India}}

\begin{document}

\title{Towards the quark mass dependence of $T_{cc}^+$ from lattice QCD}

\author{Sara Collins\orcidlink{0000-0003-0979-7602}}
\email{sara.collins@ur.de}
\reg

\author{Alexey Nefediev\orcidlink{0000-0002-9988-9430}}\email{Alexey.Nefediev@ijs.si}
\ijs \lis

\author{M. Padmanath\orcidlink{0000-0001-6877-7578}}
\email{padmanath@imsc.res.in}
\imsc

\author{Sasa Prelovsek\orcidlink{0000-0002-7496-6188}}
\email{sasa.prelovsek@ijs.si}
\ijs \ulj

\preprint{IMSc/24/01}

\begin{abstract}
The $DD^*$ scattering phase shifts in the $T_{cc}^+=cc\bar{u}\bar{d}$ channel are extracted from lattice QCD for five different charm quark masses and a fixed light-quark mass corresponding to $m_\pi\simeq 280$~MeV. The phase shifts are analysed employing two approaches: effective range expansion and Lippmann--Schwinger equation derived in the effective field theory. In the latter case, the results imply an attraction at short range parametrised by contact terms and a slight repulsion at long range mediated by one-pion exchange with $m_\pi >m_{D^*}-m_D$. The poles in the amplitude across the complex energy plane are extracted and their trajectories are discussed as the charm quark mass is varied.
Two complex conjugate poles corresponding to a resonance below threshold are found for $m_c$ close to the physical value. They turn into a pair of virtual states at the largest $m_c$ studied. With further increasing $m_c$, one virtual pole representing $T_{cc}^+$ is expected to move towards the two-body threshold and turn into a bound state. The light-quark mass dependence of the $T_{cc}^+$ pole is briefly discussed using the data on $DD^*$ scattering from other lattice collaborations.
\end{abstract}

\maketitle

\section{Introduction}

The past two decades have witnessed the arrival of a wealth of new experimental information on hadronic states with properties at odds with quark model predictions. Such states, conventionally referred to as exotic, are the focus of many theoretical investigations---see, for example,
Refs.~\cite{Esposito:2014rxa,Lebed:2016hpi,Chen:2016qju,Guo:2017jvc,Kalashnikova:2018vkv,Yamaguchi:2019vea,Brambilla:2019esw,Guo:2019twa,Chen:2022asf}. For an overview of the experimental situation and theoretical approaches see Ref.~\cite{Brambilla:2019esw}.
Most of the exotic states that have been discovered contain a heavy quark-antiquark pair ($c\bar{c}$ or $b\bar{b}$), however, experimental signatures of completely new types of exotic states with heavy quarks have recently been detected. This includes the fully charmed tetraquarks seen in the di-charmonium production spectrum \cite{LHCb:2020bwg,CMS:2023owd,ATLAS:2023bft} and the doubly charmed tetraquark $\Tcc$ \cite{LHCb:2021vvq,LHCb:2021auc} observed in the proton-proton collisions at the LHC. The latter, being the first representative of a potentially rich family of exotic hadrons with two heavy quarks, has attracted a lot of attention in the hadronic community. A remarkable feature of $\Tcc$ is its proximity to the $D^{*+}D^0$ and $D^{*0}D^+$ thresholds and its quantum numbers $J^P=1^+$ \cite{LHCb:2021vvq}, consistent with $S$-wave scattering in a pseudoscalar-vector mesons system. These two facts taken together suggest that its wave function is dominated by a long-range molecular component \cite{Guo:2017jvc}. In this scenario, the $\Tcc$ peak is associated with a shallow quasibound\footnote{We call ``quasibound'' pole, a pole located below the two-body $DD^*$ threshold that would reside on the real axis if $D^*$ were stable. We call ``binding energy'' the difference between the real part of the $\Tcc$ pole position in the energy complex plane and the $D^{*+}D^0$ threshold while ``width'' is defined as twice the imaginary part of the pole position.} pole with a binding energy of around 350~keV and a width of around 60~keV (see, for example, Refs.~\cite{LHCb:2021auc,Du:2021zzh,Dai:2023kwv}). It is instructive to note that, for the physical pion mass, the $\Tcc$ width is found to be very sensitive to the three-body effects related to the pion exchange between $D^{(*)}$ mesons. In particular,
the deviation from the above value of the extracted imaginary part of the $\Tcc$ pole may come to about 30\%, if the three-body unitarity is violated~\cite{Du:2021zzh}. Hence, setting up a meaningful low-energy expansion for the $DD^*$ scattering amplitude also requires the three-body effects to be taken into account consistently \cite{Baru:2021ldu}.

\begin{figure}[t]
\centerline{\includegraphics[width=0.8\columnwidth]{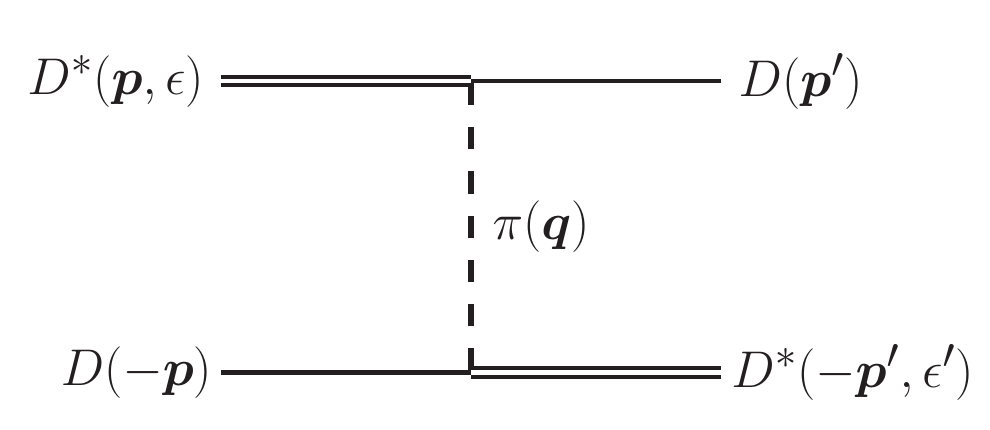}}
\caption{The Feynman diagram for the one-pion exchange interaction in a $DD^*$ system in the centre-of-mass frame. In parentheses, we indicate the momentum and (if applicable) polarisation vector of the corresponding particle.}
\label{fig:ope}
\end{figure}

Recently, the $\Tcc$ state was also studied using the lattice QCD framework in Refs.~\cite{Lyu:2023xro,Padmanath:2022cvl,Chen:2022vpo}. In Ref.~\cite{Lyu:2023xro}, the HALQCD method was employed in a simulation with $m_\pi\simeq 146$~MeV. This procedure involves extracting the $DD^*$ scattering potential, which is then used to evaluate the phase shifts above the two-body threshold. In Refs.~\cite{Padmanath:2022cvl,Chen:2022vpo}, the conventional L\"uscher's method was employed to extract the $DD^*$ phase shifts $\delta(E)$ on lattice QCD ensembles with $m_\pi\simeq 280$ and 348~MeV, respectively. All the simulations were performed in the isospin limit and for pion masses exceeding the physical value, such that the $DD\pi$ threshold lies above the $DD^*$ threshold and the decay $D^*\to D\pi$ is kinematically forbidden. The radiative decay mode $D^*\to D\gamma$ is also absent. All the above analyses assumed validity of the effective range expansion (ERE) near the $DD^*$ threshold \cite{Padmanath:2022cvl,Chen:2022vpo,Lyu:2023xro}. This rendered a virtual pole\footnote{In the given lattice settings with a stable $D^*$, this is a generic virtual pole on the real axis below the two-body $DD^*$ threshold on the second Riemann sheet with Im$(p)<0$ (here $p$ is the 3-momentum in the $DD^*$ system).} below threshold in Refs.~\cite{Padmanath:2022cvl,Lyu:2023xro}, while the authors of Ref.~\cite{Chen:2022vpo} refrained from extracting the pole position. The validity of the effective range expansion in these unphysically heavy light quark mass set-ups was questioned in Refs.~\cite{Du:2023hlu,Meng:2023bmz}, where the authors presented a re-analysis of the finite volume spectrum determined in Ref.~\cite{Padmanath:2022cvl} without assuming effective range expansion. In particular, it was pointed out that possible pion exchange interactions between $D^{(*)}$ mesons \emph{inter alia} imply the presence of the so-called left-hand cuts in the scattering amplitude with the most relevant branch point related to one-pion exchange (see Fig.~\ref{fig:ope}) lying slightly below the $DD^*$ threshold, if $m_\pi > m_{D^*}-m_D$.

The present lattice simulation aims at exploring the heavy-quark mass dependence of $DD^*$ scattering in the $\Tcc$ channel. This has not been studied before. In particular, the simulations in Ref.~\cite{Chen:2022vpo,Lyu:2023xro} were performed at the physical charm quark mass and the authors of Ref.~\cite{Padmanath:2022cvl} considered only two masses very close to the physical point. In this paper, we extract the $DD^*$ scattering amplitude for the $cc\bar u\bar d$ system with $I=0$, $J^P=1^+$, and $m_\pi\simeq 280$~MeV for five heavy-quark masses $m_c$, corresponding to $D$-meson masses in the range $m_D\simeq 1.7-2.4$~GeV. It is, therefore, an extension of the simulation in Ref.~\cite{Padmanath:2022cvl} to three additional heavier values of $m_c$. We aim to address the hypothesis whether the pole representing $\Tcc$ is approaching a bound state pole as the heavy-quark mass is increased, as expected from the theory predictions of a tightly bound $\Tbb$ with respect to the $BB^*$ threshold---see, for example, the references contained in a recent work \cite{Hudspith:2023loy}. In particular, theory predictions for $\Tbc$ with $I=0$ and $J^P=1^+$ do not yet agree on whether the state corresponds to a bound state below the $DB^*$ threshold \cite{Francis:2018jyb,Padmanath:2023rdu,Alexandrou:2023cqg} or not \cite{Hudspith:2020tdf,Meinel:2022lzo}.  The more recent simulations with either more relevant operators \cite{Alexandrou:2023cqg} or continuum and chiral extrapolation \cite{Padmanath:2023rdu} suggest that a shallow bound state is more likely.  We remark that the reduced mass of the physical $DB^*$ system, $\muD(DB^*) \simeq 1.38$~GeV, lies only slightly above the highest reduced mass of the $DD^*$ system, $\muD(DD^*) \simeq 1.23$~GeV in our study. Thus the present results may shed light on the nature of the $\Tbc$, although such conclusions may be influenced by cut-off and other unquantified effects which lie beyond the scope of the present work.

\begin{figure}[t!]
\centerline{\includegraphics[width=1.07\columnwidth]{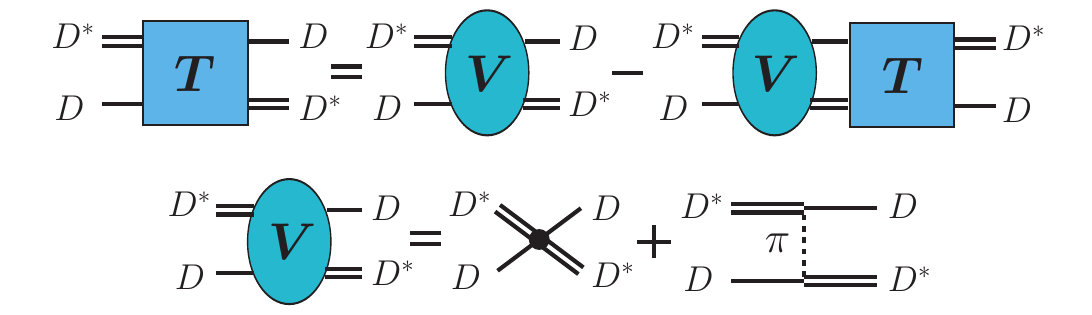}}
\caption{Graphical representation of the Lippmann--Schwinger equation for the $DD^*$ scattering amplitude that arises in the effective field theory framework.}
\label{fig:eft}
\end{figure}

This study is an improvement with respect to our previous lattice work \cite{Padmanath:2022cvl} also in terms of the extraction of energy dependence of the scattering amplitude. In addition to the effective range fit, we employ a more sophisticated approach along the lines of the analysis presented in Ref.~\cite{Du:2023hlu}. In the effective field theory (EFT) approach of Ref.~\cite{Du:2023hlu}, the interaction in the $DD^*$ system incorporates the one-pion exchange in Fig.~\ref{fig:ope} and a short-range contact potential to order ${\cal O}(p^2)$. The energy dependence of the $DD^*$ scattering amplitude is studied by solving a Lippmann--Schwinger equation, illustrated in Fig.~\ref{fig:eft}, and the coefficients of the contact potential are determined by fitting to the lattice data. The $\Tcc$ state is associated with the near-threshold pole of the resulting amplitude. It is important to emphasise, however, that the one-pion exchange gives rise to the left-hand cut \cite{Frautschi:1960qzm,Oller:2019rej} and a pole in the function $\pcot$ (with $\delta_0$ denoting the $S$-wave scattering phase), both residing just below the two-body $DD^*$ threshold. For now, the fitted scattering amplitude in the lowest partial wave is extracted from the lattice eigen-energies using L\"uscher's relation. Strictly speaking, this relation applies only above the left-hand cut and calls for modifications below it as proposed, for example, in Refs.~\cite{Raposo:2023oru,Raposo:2023nex,Dawid:2023jrj,Meng:2023bmz,Hansen:2024ffk}. Thus, like in Ref.~\cite{Du:2023hlu}, our present analysis based on the effective field theory refrains from considering the lattice levels residing below the left-hand cut branch point. 
Finally, the light-quark mass dependence of the $\Tcc$ pole is examined qualitatively by comparing the scattering amplitudes and poles obtained from the available lattice data at $m_\pi\simeq 146$~MeV \cite{Lyu:2023xro}, 280~MeV \cite{Padmanath:2022cvl}, and 348~MeV \cite{Chen:2022vpo} using the effective field theory approach outlined above. 

The paper is organised as follows. Section~\ref{sec:ope} is devoted to a pedagogical review of the pion exchange interaction in the $DD^*$ system. In Sec.~\ref{sec:lat}, we introduce the lattice data utilised in this work. The results assuming the effective range expansion are summarised in Sec.~\ref{sec:results-ere}. Section~\ref{sec:eft} contains details of the effective field theory framework employed in the lattice data analysis. We present the results obtained from the effective field theory framework in Sec.~\ref{sec:results} before concluding in Sec.~\ref{sec:conclusions}. We provide additional details and auxiliary information in several appendices. 

\begin{table*}[t!]
\begin{ruledtabular}
\begin{tabular}{cccccc}
Set (different $m_c$, all $m_\pi \simeq 280$~MeV) & 1 & 2 & 3 & 4 & 5\\
\hline
$m_D$~[GeV]& 1.762(1)&1.927(1) &2.064(2) &2.191(2) &2.415(2) \\
$m_{D^*}$~[GeV]&1.898(2) &2.049(2) &2.176(2) &2.294(2) &2.506(2) \\
$\Eth =m_D+m_{D^*}$~[GeV]& 3.660(3) & 3.976(3)& 4.240(3) &4.485(3) & 4.922(3) \\
$\mbar =\frac{1}{4}(m_D+3m_{D^*})$~[GeV]&1.864(2) & 2.019(2) & 2.148(2) & 2.269(2) & 2.484(2) \\
$\muD=(m_D^{-1}+m_{D^*}^{-1})^{-1} $~[GeV]& 0.914(1) & 0.993(1) & 1.059(1)& 1.121(1)& 1.230(1) \\
$\kappa_c$& 0.12522& 0.12315& 0.12133145& 0.11956530&0.11627907 \\
\hline
\hline
\multicolumn{6}{c}{Effective range expansion}\\
 \hline
 $a_0$~[fm] & 1.3($^{+1.0}_{-0.6}$) & 1.4($^{+1.3}_{-0.7}$) & 1.5($^{+1.6}_{-0.7}$) & 1.7($^{+2.0}_{-0.9}$) & 3.2($^{~~\infty}_{-2.6}$)\\
 $r_0$~[fm] & 1.12($^{+0.24}_{-0.29}$) & 1.0($^{+0.2}_{-0.3}$) & 0.93($^{+0.25}_{-0.25}$) & 0.87($^{+0.23}_{-0.31}$) & 0.85($^{+0.25}_{-0.27}$)\\
\vphantom{$\Bigl($} pole: $\Ep-\Eth $~[MeV] & -7($^{+5}_{-13}$) & -6($^{+4}_{-13}$)& -5($^{+4}_{-8}$)& -4($^{+3}_{-12}$) & -1.2($^{+1.2}_{-6.7}$)\\
 \hline
\hline
\multicolumn{6}{c}{Effective field theory}\\
 \hline
 $c_0$~[GeV$^{-2}$] &$-5.10\pm 0.26$ &$-4.66\pm 0.21$ & $-4.42\pm 0.50$& $-4.27\pm 0.73$& $-5.57\pm 1.92$\\
 $c_2$~[GeV$^{-4}$] &$7.64\pm 0.77$ &$6.38\pm 0.63$ &$5.57\pm 1.49$ &$4.99\pm 2.20$ & $7.28\pm 6.03$\\
 pole$_1$: $\Ep-\Eth $~[MeV] & $-8.3 - 9.1~i $& $-6.9 - 8.3~i $&$-5.9 - 7.3~i $ & $-5.0 - 6.2~i $&$-1.2 $\\
 & \scalebox{0.8}{$\quad\pm(0.7+0.8~i)$}& \scalebox{0.8}{$\quad\pm(0.5+0.8~i)$} &\scalebox{0.8}{$\quad\pm(1.1+2.0i)$} & \scalebox{0.8}{$\quad\pm(1.4+3.2~i)$}&\scalebox{0.8}{$\quad\pm 0.3$}\\
pole$_2$: $\Ep-\Eth $~[MeV] & $-8.3 + 9.0~i $& $-6.9 + 8.3~i $&$-5.9 + 7.3~i $ & $-5.0 + 6.2~i $&$-5.8 $\\
 & \scalebox{0.8}{$\quad\pm(0.7+0.8~i)$} & \scalebox{0.8}{$\quad\pm(0.5+0.8~i)$}&\scalebox{0.8}{$\quad\pm(1.1+2.0~i)$} & \scalebox{0.8}{$\quad\pm(1.4+3.2~i)$} &\scalebox{0.8}{$\quad\pm 0.3$}
 \\
 $\Elhc-\Eth $~[MeV]& -8.17(7)& -7.98(5) & -7.76(5) & -7.54(4) & -7.12(4) \\
 $(\plhc/\Eth )^2\times 10^4$ &-11.1(1)&-10.03(8)&-9.16(6)&-8.41(5)&-7.23(4) 
 \end{tabular}
\end{ruledtabular}
\caption{Results for the $S$-wave $DD^*$ scattering for five heavy-quark masses at fixed $m_\pi \simeq 280$~MeV: Sets 1 and 2 have already been analysed in Ref.~\cite{Padmanath:2022cvl}; Set 2 is closest to the physical charm quark mass. Top: masses of the scattered mesons $D$ and $D^*$ and the employed hopping parameter $\kappa_c$, which is related to the bare charm quark mass. Middle: results assuming the effective range expansion. 
The extrapolation of the effective-range-expansion-based fits to the below-threshold region is questionable due to the presence of a nearby left-hand cut and hence the pole positions should only be taken in the spirit of a consistency check with results from the previous work presented in Ref.~\cite{Padmanath:2022cvl}. Bottom: improved results based on the effective field theory taking into account the effects of the pion exchange. In the latter case, for Sets 1-4, pole$_1$ and pole$_2$ are complex conjugate poles that correspond to the $\Tcc$ as a resonance; for Set 5, the $\Tcc$ is associated with the virtual pole$_1$ while pole$_2$ (marked with asterisk) corresponds to another, more remote virtual pole. }
\label{tab:results}
\end{table*}

\section{Pion exchange and left-hand cuts}
\label{sec:ope}

To describe the $u$-channel pion exchange
in Fig.~\ref{fig:ope},
we notice that the pion emission and absorption vertices can be derived from the lowest-order nonrelativistic interaction Lagrangian~\cite{Fleming:2007rp,Hu:2005gf},
\be
{\cal L}=\frac{g_c}{2f_\pi}\left({\bm D}^{*\dagger}\cdot{\bm\nabla}\pi^a\tau^a D
+D^\dagger \tau^a{\bm\nabla}\pi^{a\dagger}\cdot{\bm D}^*\right),
\label{lagfull}
\ee
where 
$$
\tau^a\pi^a=
\begin{pmatrix}
\pi^0 & \sqrt{2}\pi^+\\
\sqrt{2}\pi^- & -\pi^0
\end{pmatrix}
$$
and $f_\pi=92.2$~MeV and $g_c=0.57$ are the pion decay constant  and the $DD^*\pi$ coupling, respectively.\footnote{Here we quote the values of $f_\pi$ and $g_c$ relevant for the physical pion mass; the values actually employed in the calculations will be discussed below.}
Lagrangian \eqref{lagfull} renders the one-pion exchange potential
\be
V_\pi(\vep,\vep')
=3\left(\frac{g_c}{2f_\pi}
\right)^2
\frac{(\veep\cdot \veq)(\veq\cdot {\veep^{\prime *}})}{u-m_\pi^2}, \label{VOPE}
\ee
where the momenta $\vep^{(\prime)}$ and polarisation vectors $\veep^{(\prime)}$ are defined in Fig.~\ref{fig:ope}, the isospin factor for the isoscalar $DD^*$ system is taken into account explicitly, $\veq=\vep'-\vep$ and
$$
u-m_\pi^2=q^2-m_\pi^2=-\veq^2-\left(m_\pi^2-q_0^2\right).
$$

The potential in Eq.~(\ref{VOPE}) is used in our actual calculations (see Eq.~(\ref{Swave}) and below), while here we discuss its simplified version in order to introduce several theoretical concepts. In particular, if the $D^{(*)}$-meson recoil terms are neglected, then 
$$
q_0=E_{D^*}-E_D\approx m_{D^*}-m_D,
$$
and the denominator of the pion propagator can be re-written in the form
\be
u-m_\pi^2\approx -(\veq^2+\mu_\pi^2),
\ee
where the effective pion mass parameter \cite{Tornqvist:1993ng,Swanson:2003tb,Suzuki:2005ha},
\be
\mu_\pi^2=m_\pi^2-(m_{D^*}-m_D)^2,
\label{mupi}
\ee
defines the long-range behaviour of one-pion exchange. In order to see it explicitly, we first re-write the tensor $q_iq_j$ appearing in the numerator of the expression in Eq.~(\ref{VOPE}) as
\be
q_iq_j=\frac13\delta_{ij}\veq^2+\left(q_iq_j-\frac13\delta_{ij}\veq^2\right)
\label{cent}
\ee
and then consider the central part of the potential (\ref{VOPE}) that corresponds to retaining only the first term in Eq.~(\ref{cent}),
\be
V_\pi(\vep,\vep')\Longrightarrow (\veep\cdot \veep^{\prime *})V_\pi^{\rm cent}(\veq),
\ee
with
\be
V_\pi^{\rm cent}(\veq)=\left(\frac{g_c}{2f_\pi}\right)^2
\frac{\veq^2}{u-m_\pi^2}
=\frac{g_c^2}{4f_\pi^2}\left(-1+\frac{\mu_\pi^2}{\veq^2+\mu_\pi^2}\right).
\label{VOPE0}
\ee
The first term in parentheses on the right-hand side of Eq.~(\ref{VOPE0}) describes the attractive short-range part of the one-pion exchange interaction, which is proportional to $\delta^{(3)}(\ver)$ in coordinate space. The second term in parentheses corresponds to the long-range part that is fully defined by the known coupling $g_c$ and the effective mass parameter from Eq.~(\ref{mupi}) rather than the pion mass alone. The physical (tagged as $\0$) masses correspond to
$m_\pi\0 <m_{D^*}\0-m_D\0$ and $\oo{\mu_\pi}<0$. However, in all available lattice simulations, $m_\pi^{\rm lat}>m_{D^*}^{\rm lat}-m_D^{\rm lat}$, such that $(\mu_\pi^{\rm lat})^2>0$.\footnote{In what follows the superscript $^{\rm lat}$ will be omitted and, unless explicitly stated, all quantities are the lattice ones.} Therefore, the behaviour of the one-pion exchange potential at large distances in $r$-space changes dramatically when the pion mass changes from its physical value to an unphysical scenario with $m_\pi>m_{D^*}-m_D$. Since the long-range part in the potential is proportional to $\mu_\pi^2$ (see also Eq.~(\ref{VOPEr})), in the latter case, it has the opposite sign to that for the physical $\Tcc$, which translates to a slight repulsion at intermediate and long range compared to a purely attractive behaviour in the physical system.\footnote{In the latter case, the very notion of a potential needs to be treated with caution since the Fourier transform of Eq.~(\ref{VOPE0}) results in an oscillating function of $r$ that has an imaginary part; we, therefore, refer to the potential in the same sense as in Refs.~\cite{Suzuki:2005ha,Liu:2008fh,Thomas:2008ja}.}

Some further properties of the simplified potential in Eq.~(\ref{VOPE0}) are discussed in Appendix~\ref{app:pot}. However, we emphasise that this simplified potential is never used in any calculations in this work, and its 
discussion is solely aimed at motivating the repulsive nature of the long range interaction in the lattice set-up. For a detailed study of the one-pion exchange potential in heavy-meson systems in coordinate space, {\em c.f.} Ref.~\cite{Tornqvist:1993ng,Swanson:2003tb,Suzuki:2005ha,Liu:2008fh,Thomas:2008ja}.

Performing the $S$-wave projection of the potential in Eq.~(\ref{VOPE}) (see, for example, Appendix~B of Ref.~\cite{Baru:2019xnh} for details),
\begin{align}
V_\pi^S(p,p')&=\frac1{2J+1}\int\frac{d \Omega_p}{4\pi}\frac{d \Omega_{p'}}{4\pi} \sum_{\sigma,\sigma'}(\veep^*\cdot\veep')V_\pi(\bm p,\bm p'),\nonumber\\
&=\frac{g_c^2}{8 f_\pi^2} \int_{-1}^1 \frac{\veq^2}{u-m_\pi^2}d\cos\theta,\label{Swave}
\end{align}
where $J=1$ is the total spin of $\Tcc$, $\sigma^{(\prime)}$ are the polarisations of the initial(final)-state $D^*$ meson, $\theta$ is the angle between the 3-momenta $\vep$ and $\vep'$ and $\veep$ and $\veep'$ are the polarisation vectors of the $D^*$ meson in the initial and final state, respectively, that also play the role of $S$-wave projectors for the $DD^*$ system with $J^P=1^+$. 

For $|\vec p|=|\vec p^\prime|=p$, the $S$-wave projected potential (\ref{Swave}) reads
\begin{align}
V^{S}_{\pi}(p,p)=\frac{g_c^2}{8 f_\pi^2} \int_{-1}^1
 \frac{ 2p^2(1- \cos\theta)\;d\cos\theta}{q_0^2- 2p^2(1- \cos\theta)-m_\pi^2}.
\label{VOPESS}
\end{align}
The angular integration is trivially performed giving a logarithmic function,
$$
V^{S}_{\pi}(p,p)=\frac{g_c^2}{4f_\pi^2}\left[\frac{m_\pi^2-q_0^2}{4p^2}\ln \left(1+\frac{4p^2}{m_\pi^2-q_0^2}\right)-1\right],
$$
that introduces an infinite set of Riemann sheets attached to each other at the branch point located at
\be
\plhc^2 =\frac14(q_0^2-m_\pi^2)\simeq -\frac14\mu_\pi^2 <0,
\label{lhc}
\ee
for $q_0\simeq m_{D^*}-m_D$. 
This logarithmic cut in the potential, that starts at the branch point $\plhc^2$ and is conventionally chosen to run to $-\infty$ along the real axis, is known as a {\it left-hand cut}. In the lattice settings here and in all previous studies, this cut starts very close to the $DD^*$ threshold and needs to be taken into account in the calculations, as argued in Ref.~\cite{Du:2023hlu}.

\section{Lattice simulation and eigen-energies}
\label{sec:lat}

The eigen-energies of the $cc\bar{u}\bar{d}$ system and the resulting $S$-wave $DD^*$ scattering phase shifts are extracted on two $N_f=2+1$ lattice ensembles with $a=0.08636(98)(40)$~fm and $m_\pi=280(3)$~MeV with different spatial extents, $N_L=24$ and 32, generated by the Coordinated Lattice Simulations (CLS) consortium~\cite{Bruno:2014jqa,Bali:2016umi}.\footnote{The corresponding CLS ensembles are labelled U101 and H105, respectively.} The nonperturbatively improved Wilson-clover action is employed for all quark fields. Five values of the charm quark hopping parameter $\kappa_c$ are realised to explore the heavy-quark mass dependence of the finite volume spectrum. The corresponding masses of the $D$ and $D^*$ mesons, their spin average $\mbar=(m_D+3m_{D^*})/4$, the $DD^*$ threshold $\Eth = (m_D+m_{D^*})$, and other relevant quantities are given in Table \ref{tab:results}. Sets 1 and 2 were studied already in our previous simulation~\cite{Padmanath:2022cvl}, where Set 2 is closest to the physical charm quark mass.

We employ the same set of interpolators $D(\vec p_1)D^*(\vec p_2)$ and $D^*(\vec p_1)D^*(\vec p_2)$ that resemble two-meson scattering channels as in Ref.~\cite{Padmanath:2022cvl}. Interpolators with diquark-antidiquark colour structure $[cc][\bar u\bar d]$ are not employed in the present simulation.
The irreducible representations 
for total momenta $|\vec P|=|\vec p_1+\vec p_2|=0$, $1\frac{2\pi}{L}$, that are used in the analysis,
are listed in Table \ref{tab:irreps}. The correlation matrix is evaluated using the distillation method \cite{HadronSpectrum:2009krc,Piemonte:2019cbi} and eigen-energies are extracted by solving the generalised eigenvalue problem \cite{Michael:1985ne}.

\begin{figure*}[t!]
\includegraphics[height=0.5\textwidth,width=0.99\textwidth]{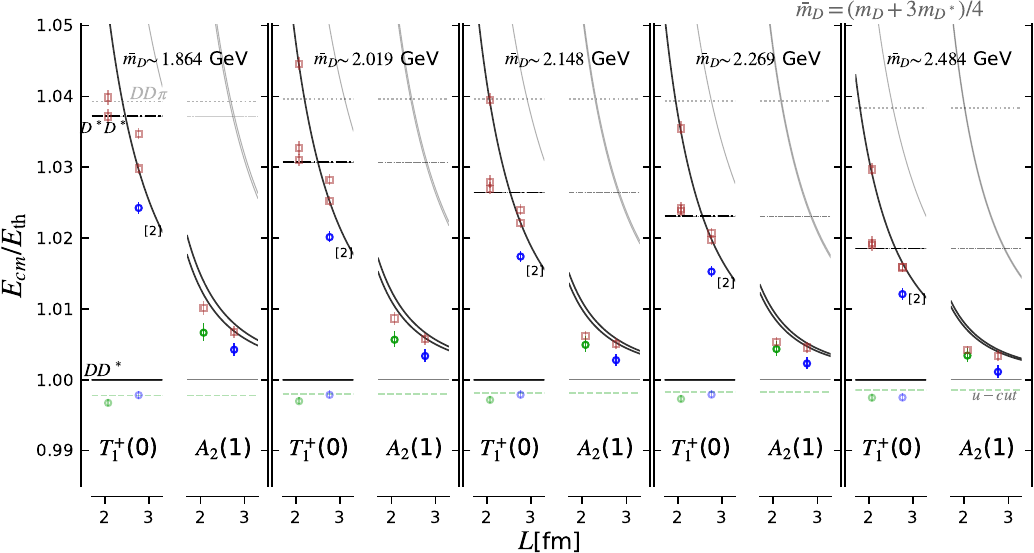}
\caption{The centre-of-momentum energy $E_{\rm cm}=(E^2-\vec P^2)^{1/2}$ of the $cc\bar u\bar d$ system normalised by $\Eth \equiv m_D+m_{D^*}$ for five heavy-quark masses and various irreps (listed in Table \ref{tab:irreps}). The lattice eigen-energies are shown by the empty symbols: the $S$-wave $DD^*$ scattering amplitudes are extracted using the dark shaded blue and green circles. The noninteracting $DD^*$ energies in Eq.~(\ref{Eni}) in the centre-of-momentum frame are shown by curves.}
\label{fig:En}
\end{figure*}

The eigen-energies of the $cc\bar u \bar d$ system for the five heavy-quark masses and total momenta $|\vec P|=|\vec p_1+\vec p_2|=0$, $1\frac{2\pi}{L}$ are presented in Fig.~\ref{fig:En}. The most notable difference between the spectra is that the separation between the $DD^*$ and $D^*D^*$ thresholds decreases as the heavy-quark mass increases, as expected from the dependence $m_{D^*}-m_{D}\propto 1/m_c$
(see, for example, Ref.~\cite{Manohar:2000dt} for
details of Heavy Quark Effective Theory expectations). Apart from that, all the spectra exhibit similar features. Note that most of the energies are below the noninteracting $DD^*$ levels,
\begin{equation}
\label{Eni}
E^{\mathrm{ni}}= E_{D(\vec p_1)} +E_{D^*(\vec p_2)} ~,\ \vec p_i=\vec n_i\frac{2\pi}{L}, \ \vec n_i\in N^3,
\end{equation}
implying that we observe an attractive interaction between $D$ and $D^*$. The energy shifts are of a similar order of magnitude for the different quark masses. The eigen-states that couple dominantly to the interpolators $(DD^*)_{l=2}$ are not shifted significantly, while those related to $(DD^*)_{l=0}$ are shifted down and are employed to extract the scattering amplitude. The levels that have dominant overlaps with $D^*D^*$ operators are observed to show nonnegligible negative energy shifts, and hence we limit our fits to the elastic region well below the $D^*D^*$ threshold. We also observe nonnegligible energy shifts in the levels related to $(DD^*)_{l=1}$ operators, consistent with the observations from our previous work \cite{Padmanath:2022cvl}. However, we observe that the $S$-wave parameters extracted from a pure $S$-wave fit and a combined $S$- and $P$-wave fit are consistent within errors. Additionally, our fit results based on the effective field theory approach are restricted to $S$ wave. For these reasons, 
we only present results from 
pure $S$-wave fits in this work.

\begin{table}[t!]
\begin{ruledtabular}
\begin{tabular}{ccc | c|c }
$\vec P$ & Symmetry & Irrep$^P$ & $J^P$ & $l$ \\
\hline
$(0,0,0)$ & $O_h$ & $T_1^+$ & $1^+$ & $0,2$ \\
$(0,0,1)\frac{2\pi}{L}$ & $\mathrm{Dic}_4$ & $A_2$ & $0^-,1^+,2^-$ & 0,1,2
\end{tabular}
\end{ruledtabular}
\caption{The total momentum $\vec P$, spatial lattice symmetry group and irreducible representations analysed for the $cc\bar u\bar d$ system, together with $J^P$ and partial-wave $l$ of $DD^*$ scattering that contributes to each irrep (only $J,l\leqslant 2$ are listed). }
\label{tab:irreps}
\end{table}

\section{Results assuming effective range expansion}
\label{sec:results-ere}

In this section, we present the $S$-wave $DD^*$ scattering amplitude $T_0$ extracted from the eigen-energies above the $DD^*$ threshold in the elastic region using L\"uscher's formalism for two-body scattering. The energy dependence is parametrised with an effective range expansion (ERE). Truncating the effective range expansion after the first two terms, we have
\be
-\frac{2\pi}{\muD}T_0^{-1}=\pcot-ip,
\label{amplT}
\ee
where
\be
\pcot=\frac{1}{a_0}+\frac12r_0p^2.
\label{ERE}
\ee
Here $a_0$ and $r_0$ denote the scattering length and effective range, respectively, and $\muD$ denotes the $DD^*$ reduced mass---see Table~\ref{tab:results}. The same parametrisation was utilised in our previous study~\cite{Padmanath:2022cvl}, in which the analysis also considered the finite-volume levels below the $DD^*$ threshold, where left-hand cut effects could be dominant.

\subsection{Heavy-quark mass dependence}
 
\begin{figure}[t!]
\includegraphics[height=0.65\textwidth,width=0.97\columnwidth]{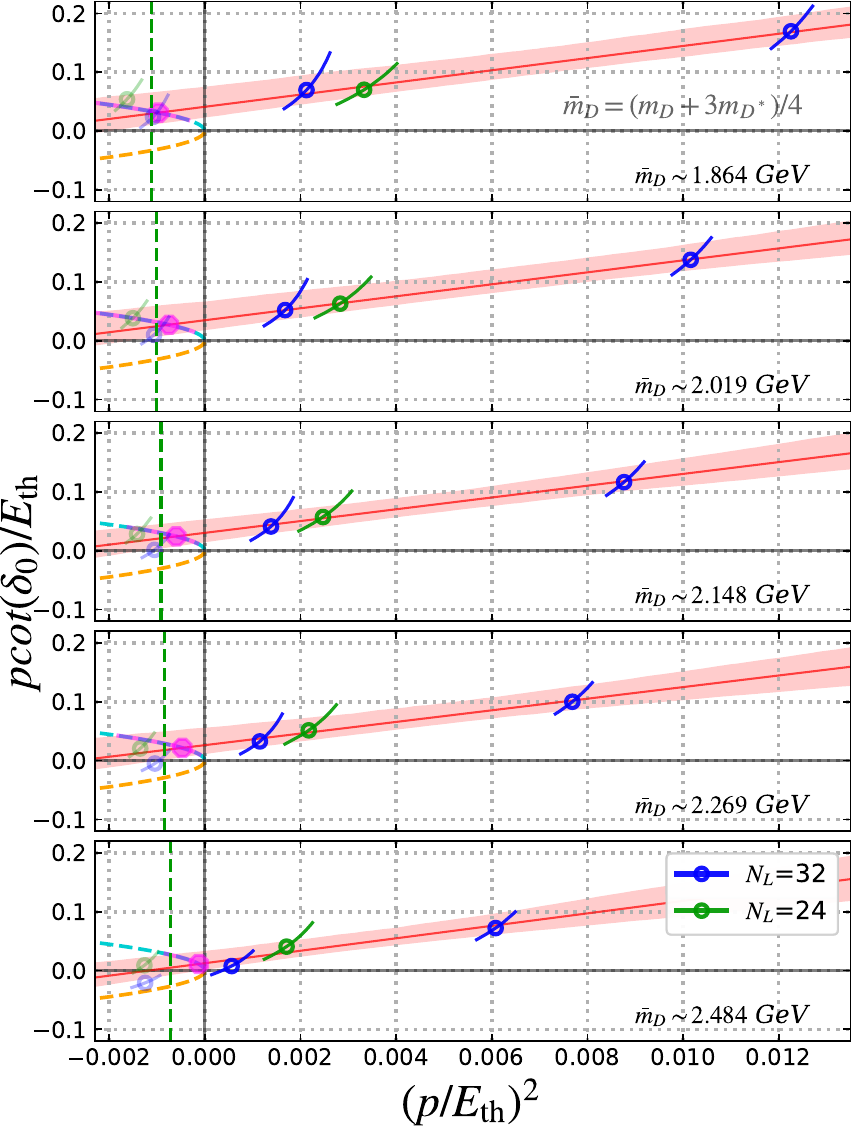}
\caption{The plots of $\pcot$ versus $p^2$ in units of $\Eth$ for the five heavy-quark masses studied. The bands represent the effective range expansion fits. The fit excludes the levels indicated by the light-shaded symbols that are near or below the left-hand cut (green vertical line).
The magenta symbols refer to the poles in the amplitude, where the effective range expansion fits cross the unitary parabola $ip$ indicated by the dashed lines. The poles are close to the left hand cut and should be taken with a grain of salt as the applicability of effective range expansion is questionable in this region.
} 
\label{fig:pcotd-ere}
\end{figure}

\begin{figure*}[t!]
\includegraphics[width=0.96\textwidth]{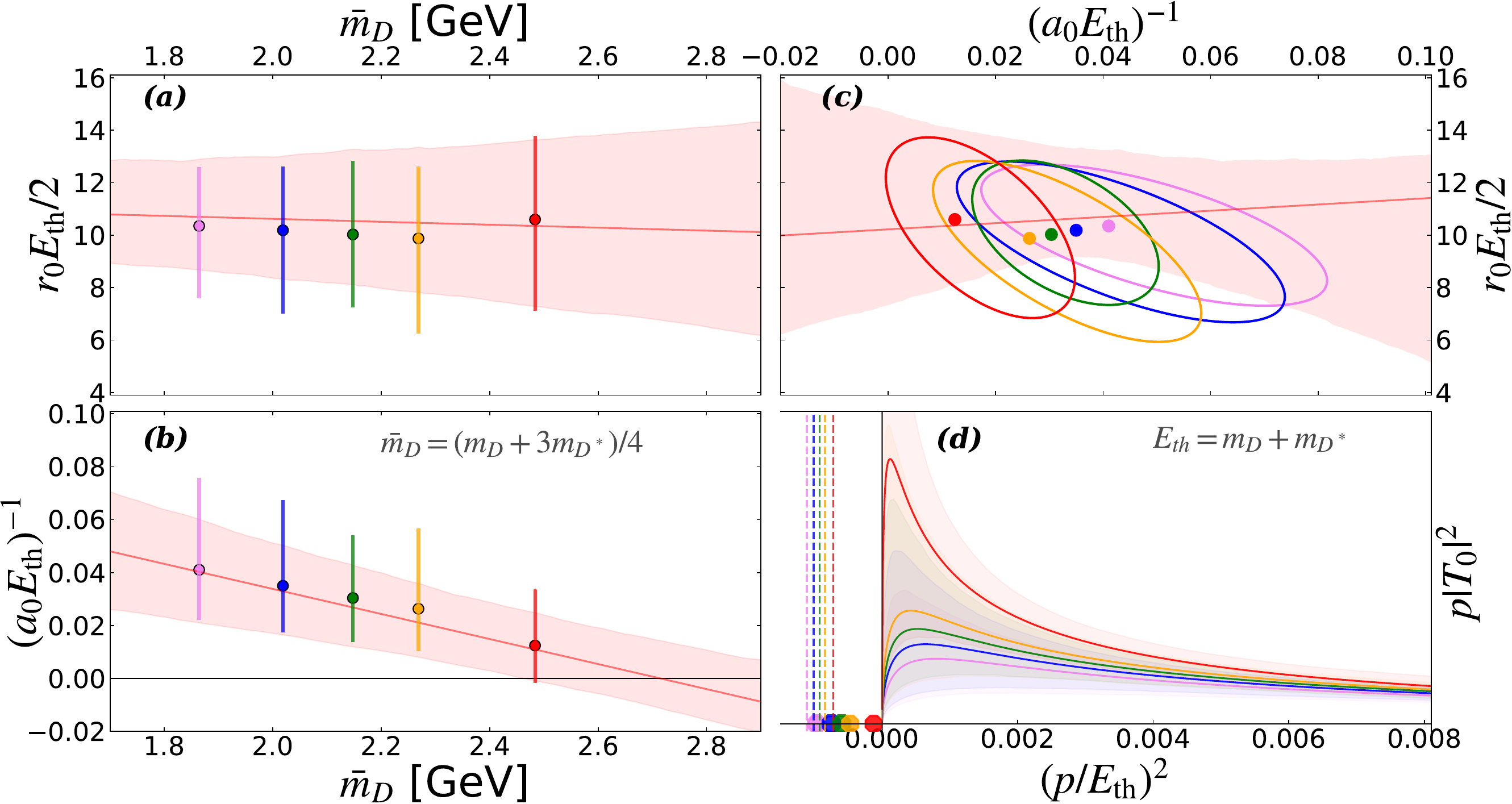}
\caption{
Heavy quark mass dependence of the effective range expansion parameters, effective range $r_0$ (a) and inverse scattering length $a_0^{-1}$ (b), in units of $\Eth$. (c) Correlated dependence of the effective range expansion parameters on the heavy-quark mass. (d) The $DD^*$ scattering rate $\propto p|T_0|^2$ above the threshold in arbitrary units. The left-hand cut branch points are shown as dashed vertical lines and the virtual
 bound poles are presented as circles.}
\label{fig:results-ere}
\end{figure*}

\begin{figure}[t!]
\includegraphics[width=0.4\textwidth]{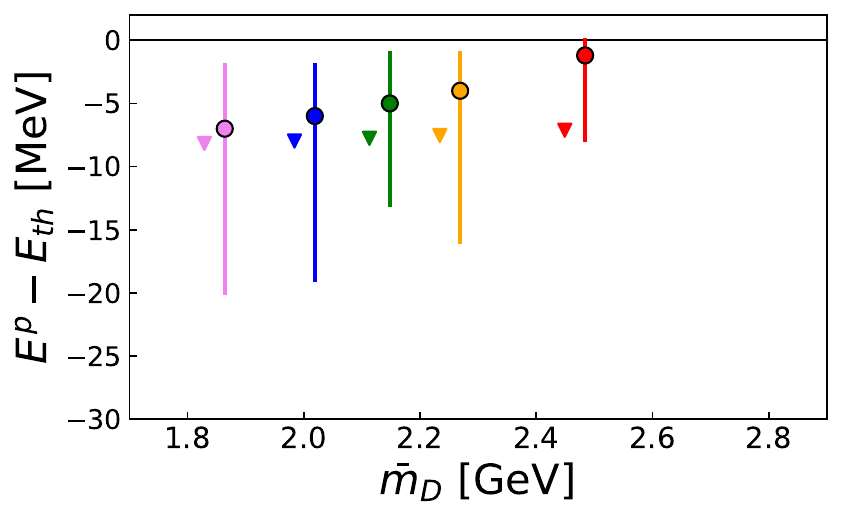}
\caption{The position of the $DD^*$ scattering amplitude pole as function of the spin-averaged mass $\mbar$ extracted using the effective range expansion approach---see Table~\ref{tab:results}. The poles presented here are extracted ignoring nonnegligble left-hand-cut-related effects and hence should only be taken in the spirit of a consistency check with results from the previous work presented in Ref.~\cite{Padmanath:2022cvl}. The position of the left-hand cut is indicated by triangles. }
\label{fig:results-ere2}
\end{figure}

A simultaneous fit to the levels indicated with dark shaded blue and green circles in Fig.~\ref{fig:En} renders the scattering amplitude for ($J=1^+$, $l=0$). The momentum-squared or equivalently the energy dependence of $\pcot$ for the different heavy-quark masses is shown in Fig.~\ref{fig:pcotd-ere}.
For all heavy-quark masses, the quantity $\pcot$ is increasing with the energy, and the scattering length $a_0$ is positive, as evident from panel $\bm{(b)}$ of Fig.~\ref{fig:results-ere}. The fit results for the effective range parameters are listed in Table~\ref{tab:results}. 
Ignoring any left-hand cut effects, the effective range fits cross the unitary parabola $ip$ at the point where
\be
\frac{1}{a_0}+\frac12r_0p^2=ip.
\label{EREpole}
\ee
The solution to this relation indicates a pole in $T_0$ within the complex energy plane on the unphysical Riemann sheet with momentum $\pp$ [Re$(\pp)$=0, Im$(\pp)<0$], as indicated in Fig.~\ref{fig:pcotd-ere} by a magenta octagon for each heavy quark mass.
Therefore, for all five heavy quark masses used in this analysis, which is blind to any left-hand cut effects, the doubly charm tetraquark corresponds to a virtual pole slightly below the $DD^*$ threshold. Its binding energy is $\Ep-\Eth$, with a pole energy in the centre-of-mass frame
\be
\Ep=(m_D^2-|\pp|^2)^{1/2} +(m_{D^*}^2-|\pp|^2)^{1/2}.
\ee
The pole positions obtained are collected in Table~\ref{tab:results} and shown in Figs.~\ref{fig:results-ere}, \ref{fig:results-ere2}, and \ref{fig:polespE}. These shallow virtual poles enhance the scattering rate $\propto p|T_0|^2$ just above threshold as demonstrated in panel $\bm{(d)}$ of Fig.~\ref{fig:results-ere}.

Considering the questionable applicability of the effective range expansion below the $DD^*$ threshold in our set-up, we emphasise that the fits involved only three levels above the threshold, shown with darker shades of green and blue colours in Figs.~\ref{fig:En} and \ref{fig:pcotd-ere}, for each heavy-quark mass considered. We find that the fit results for the scattering length and effective range following this procedure are consistent with the results one obtains using all five levels, including the levels below threshold shown in lighter shades of green and blue in Figs.~\ref{fig:En} and \ref{fig:pcotd-ere}. Note that an extrapolation of the resulting estimates for $\pcot$ below the $DD^*$ threshold, where the left-hand cut effects could be dominant, is not justified for the same reasons as for including the two subthreshold lattice levels. Hence the results for the pole positions arising out of extrapolation of the effective-range-expansion-based energy dependence of the amplitude below the threshold are naturally questionable.

\begin{figure}[t!]
\includegraphics[width=0.9\columnwidth]{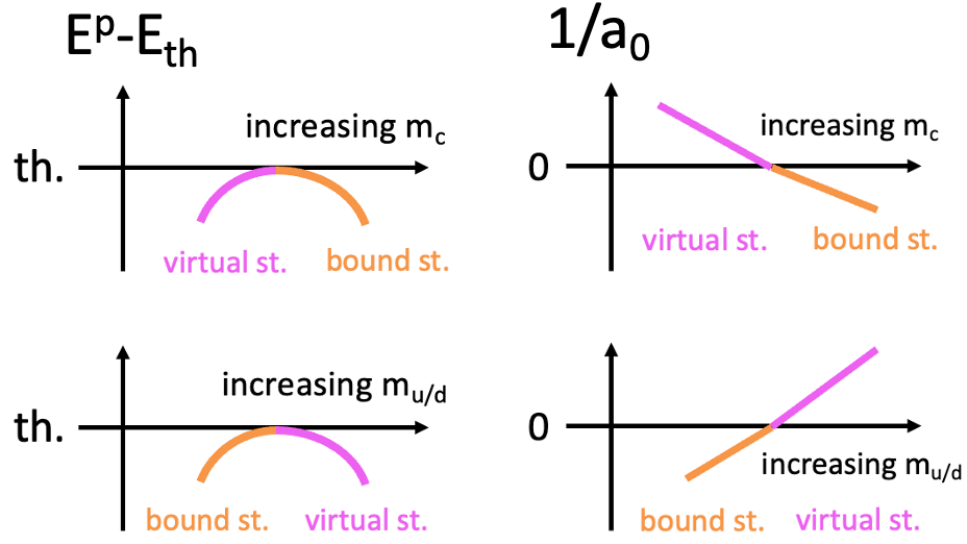}
\caption{Sketch of the quark mass dependence of the binding energy and inverse scattering length (see Eq.~\eqref{ERE}) for a (virtual) bound state in $S$ wave based on a purely attractive $DD^*$ interaction potential that is roughly $m_Q$-independent and gets stronger with decreasing $m_{u/d}$ \cite{Padmanath:2022cvl}. The quark mass dependence presented ignores left-hand-cut-related effects. }
\label{fig:ere-expectation}
\end{figure}

As the heavy-quark mass increases, the positive inverse scattering length $a_0^{-1}$ decreases, and this virtual pole approaches the threshold. For our set-up with $m_\pi\simeq 280$~MeV, the extrapolated inverse scattering length changes sign at the charm quark mass that corresponds to the critical values
\bea
\mbar^{\rm crit}(\mbox{ERE}) &=& 2.71(^{+34}_{-26})~\mbox{GeV},\nonumber \\[-2mm]
\label{critERE}\\[-2mm]
\muD^{\rm crit}(\mbox{ERE}) &=& 1.35(^{+17}_{-13}) ~\mbox{GeV}. \nonumber
\eea
This is fairly close to the values for Set~5 in Table~\ref{tab:results} which corresponds to the heaviest $c$-quark mass in our study. At this critical mass, the virtual state is expected to turn into a real bound state. The observed pattern of the heavy-quark mass dependence is sketched in Fig.~\ref{fig:ere-expectation}.
This dependence is qualitatively consistent with a purely attractive and roughly $m_Q$-independent $DD^*$ potential and a kinetic term that decreases with $m_Q$. It implies 
the existence of a strongly bound state at $m_Q\simeq m_b$ in line with the theoretical expectations of a deeply bound doubly bottomed tetraquark $\Tbb$. 

In addition to the cautionary remarks concerning the validity of the effective range expansion, we emphasise the qualitative nature of the inferences from this analysis. Cut-off and other unaccounted effects can quantitatively influence this picture. However, the investigation of these effects lies beyond the scope of the present work.

\subsection{Light-quark mass dependence}
 
The results presented in this work are generated at a fixed light quark mass $m_{u/d}$ corresponding to $m_\pi\simeq 280$~MeV. Considering also Refs. \cite{Padmanath:2022cvl,Chen:2022vpo,Lyu:2023xro}, previous lattice studies have explored the pion mass dependence of the $T_{cc}^+$ state in the range $350 \geqslant m_\pi\geqslant 146$~MeV. When utilising the effective range expansion, all these simulations lead to a virtual state and a positive $a_0^{-1}$, which increases with decreasing $m_{u/d}$ (see Fig.~1 of Ref.~\cite{Lyu:2023xro}). This is in agreement with the expectations discussed in Ref.~\cite{Padmanath:2022cvl} and the quark mass dependence illustrated in Fig.~\ref{fig:ere-expectation}, which suggests that the barely bound physical $\Tcc$ observed by LHCb becomes a virtual state at $m_\pi^{\rm lat}>m_{\pi}^{\rm ph}$.
 
\section{Effective field theory approach}
\label{sec:eft}

The wave function of the $DD^*$ system in the isoscalar channel takes the form
\be
\ket{DD^*,I=0}=\frac{1}{\sqrt{2}}(\ket{D^+D^{*0}}-\ket{D^0D^{*+}}),
\label{eq:iso_decomp}
\ee
and the $DD^*$ scattering amplitude $T$ in this system is described as illustrated in Fig.~\ref{fig:eft} and detailed below. In particular, the potential for off-shell one-pion exchange, as depicted in Fig.~\ref{fig:ope}, is provided in Eq.~\eqref{VOPE}, and its $S$-wave projection is given in Eq.~(\ref{Swave}).
To proceed, it is also pertinent to express the pion propagator $1/(u-m_\pi^2)$ in Eq.~(\ref{Swave}) as a sum of the two contributions in time-ordered perturbation theory (see, for example, Ref.~\cite{Du:2021zzh}). It gives
\begin{align}
V^{S}_{\pi}(p,p')=&\frac{g_c^2}{8 f_\pi^2} \int_{-1}^1d\cos\theta 
\left(\frac{1}{D_{DD\pi}(\vep,\vep')}\right.\label{VOPESS1}\\
&\left.+\frac{1}{D_{D^*D^*\pi}(\vep,\vep')}\right) (p^2+p'^2-2 p p' \cos\theta),\nonumber
\end{align} 
where 
\begin{align*}
D_{DD\pi}(\vep,\vep')&=-2E_{\pi}(\veq)\\
&\times\Bigl(E_D(\vep)+E_D(\vep')+E_\pi(\veq)-\sqrt{s}\Bigr),\\
D_{D^*D^*\pi}(\vep,\vep')&=-2E_{\pi}(\veq)\\
&\times\Bigl(E_{D^*}(\vep)+E_{D^*}(\vep')+E_\pi(\veq)-\sqrt{s}\Bigr),
\end{align*}
with $E_\pi(\veq)=(\veq^2+m_\pi^2)^{1/2}$, $E_{D^{(*)}}(\vep)=(\vep^2+m_{D^{(*)}}^2)^{1/2}$ and 
$\sqrt{s}=\Eth +\vep^2/(2\muD)$.

In the effective field theory framework, all additional $S$-wave interactions in the $DD^*$ system (mediated by heavy-particle exchanges) can be effectively parametrised in the form\footnote{We work in the strict heavy quark spin symmetry (HQSS) limit and take HQSS-breaking effects into account explicitly through the $D^*$-$D$ mass splitting.}
\be
V_{\rm CT}(p,p')=2c_0+2c_2(p^2+p^{\prime 2})+{\cal O}(p^4,p^{\prime 4}),
\label{VCT}
\ee
where $c_0$ and $c_2$ are unknown low-energy constants and the terms up to order ${\cal O}(p^2,p^{\prime 2})$ were explicitly retained in the low-energy expansion of the 
potential.\footnote{We use the convention of Ref.~\cite{Du:2021zzh} that differs from the convention of Ref.~\cite{Du:2023hlu} by an overall factor of 2.} We note that a fit to the lattice data renders an attractive potential at short distance.

\begin{figure}[t!]
\centerline{\includegraphics[width=0.7\columnwidth]{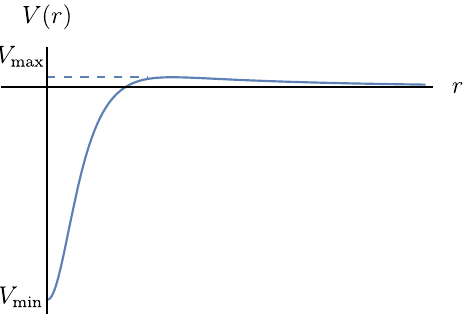}}
\caption{A simplified sketch of the central part of the total $DD^*$ potential for $\Tcc$ (see also Fig.~\ref{fig:Vr}), where the recoil terms are neglected and the mass parameter defined in Eq.~\eqref{mupi} is positive, $\mu_\pi^2>0$. A suitable regularisation is assumed at short range as introduced in Eq.~\eqref{Vreg} and explained in the text around it. This simplified presentation of the $DD^*$ interaction potential should only be regarded as qualitative and indicative.}
\label{fig:Vtot}
\end{figure}

Then the total $S$-wave interaction potential in the $DD^*$ system is a sum of the short-range part from Eq.~(\ref{VCT}) and the pion exchange term in Eq.~(\ref{VOPESS1}),
\be
V(\vep,\vep')=V_{\rm CT}(p,p')+V^S_\pi(p,p'),
\label{Vtot}
\ee
as illustrated in Fig.~\ref{fig:eft}.
This potential does not have a straightforward representation as a single-argument function in coordinate space and needs to be regularised at short range. In Appendix~\ref{app:pot} we discuss its simplified form to pinpoint some general features relevant for understanding the motion of the $DD^*$ scattering amplitude pole. In particular, a typical $V(r)$ extracted from our data has an attractive short-range part and a small barrier at larger distances as sketched in Fig.~\ref{fig:Vtot} (see also Fig.~\ref{fig:Vr} below). Note also that, in the total potential in Eq.~(\ref{Vtot}), the short-range part of the pion exchange potential (first term in parentheses in Eq.~\eqref{VOPE0}) can be kept in $V_\pi^S$ or absorbed by $c_0$ in $V_{\rm CT}$; we choose the former option. This ambiguity emphasises the fact that the short-range interaction from the contact terms and pion exchange cannot be disentangled in a scheme-independent
way---see Ref.~\cite{Baru:2015nea} for a detailed discussion.

With the total potential in Eq.~(\ref{Vtot}) at hand we formulate the Lippmann--Schwinger equation for the off-shell $DD^*$ scattering amplitude 
\be
\begin{split}
T(\vp,\vp';E)&=V(\vp,\vp')\\
&-\int\dk V(\vp,\vk)G(\vk;E)T(\vk,\vp';E),
\label{LS}
\end{split}
\ee
where the $DD^*$ loop function is taken in the form
\be
G(\vk;E)=\frac{1}{\Eth +\vk^2/(2\muD)-E+i0}.
\label{loop}
\ee
A graphical representation of the Lippmann--Schwinger equation in Eq.~\eqref{LS} with the potential in Eq.~\eqref{Vtot} is depicted in Fig.~\ref{fig:eft}. This equation is solved numerically, as detailed in Appendix~\ref{app:LSE}.

Finally, we determine and study the on-shell $DD^*$ scattering amplitude $T(E)\equiv T(\vep,\vep;E)$, where the energy and momentum are connected via the nonrelativistic dispersion relation $E=\Eth +p^2/(2\muD)$. To facilitate comparison with previous works as well as with the analysis performed using the effective range expansion, we represent $T(E)$ in the form in Eq.~(\ref{amplT}) and in what follows refer to the quantity $\pcot$.

\section{Results based on effective field theory}
\label{sec:results}

In this section, we present the lattice-extracted $S$-wave amplitudes following the effective field theory approach described in Sec.~\ref{sec:eft}, where some technical aspects are detailed in Appendix~\ref{app:LSE}. The analysis is similar in spirit to that in Ref.~\cite{Du:2023hlu}.

\subsection{Heavy-quark mass dependence }

\begin{figure}[t!]
\includegraphics[width=0.99\columnwidth]{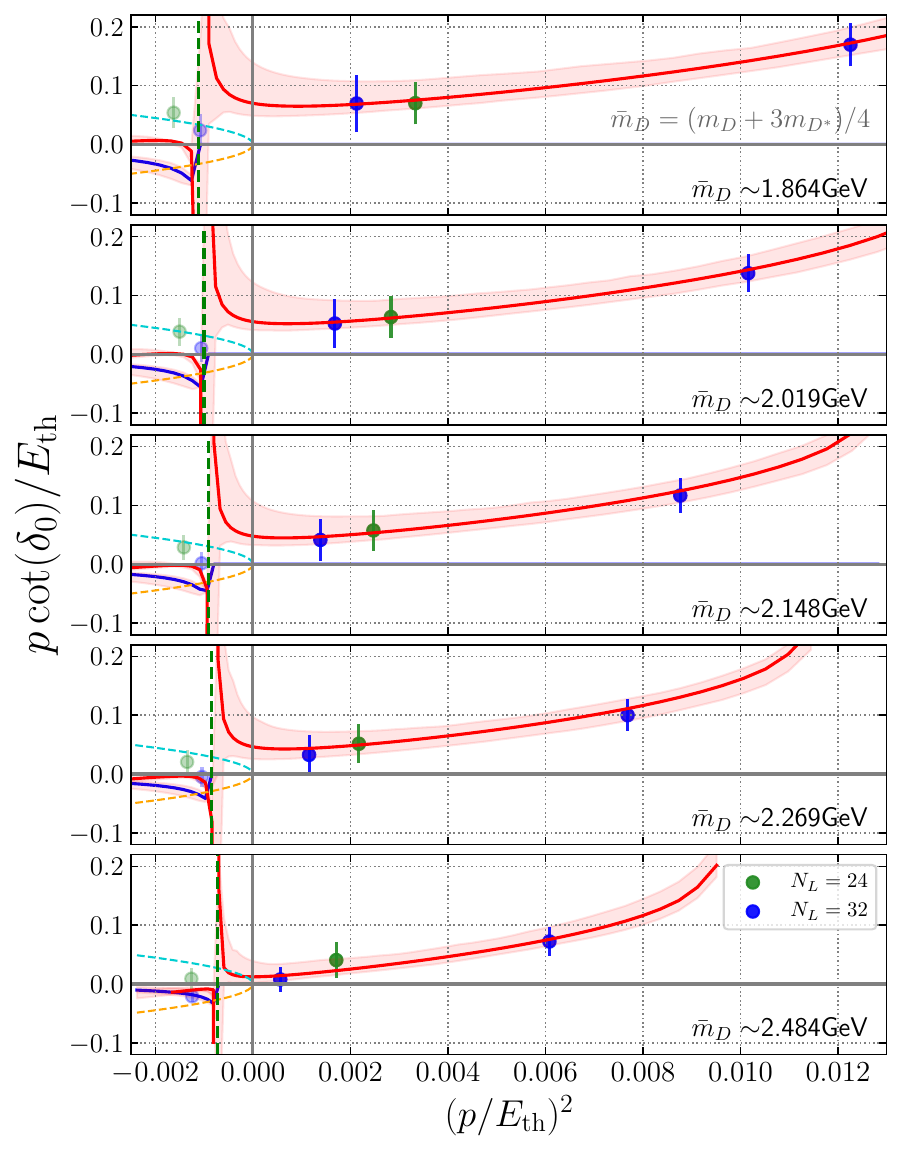}
\caption{The fits to $\pcot$ for five different values of the heavy-quark mass from the effective field theory approach. The resulting values for the fitting parameters are quoted in Table~\ref{tab:results}. }
\label{fig:pcotdelta}
\end{figure}

The lattice data for the $S$-wave scattering $\pcot$ for each of the five heavy-quark masses are fitted with the solution of the Lippmann--Schwinger equation in Eq.~(\ref{LS}), see Fig.~\ref{fig:pcotdelta}. The two fit parameters are the contact terms $c_0$ and $c_2$ introduced in Eq.~(\ref{VCT}). Only the three data points lying above the left-hand cut branch point (indicated by the green dashed vertical line in Fig.~\ref{fig:pcotdelta}) are included in the fit.\footnote{ The fit minimises $\chi^2=r_i\left(\mathrm{Cov}^{-1}\right)_{ij}r_j
$, where Cov is the covariance matrix and the residual for the $i$-th level is $r_i=(p\cot\delta_0)^{\rm lat}_i-(p\cot\delta_0)^{\rm EFT}_i$.} 

\begin{figure}[t!]
\centerline{
\includegraphics[width=0.85\columnwidth]{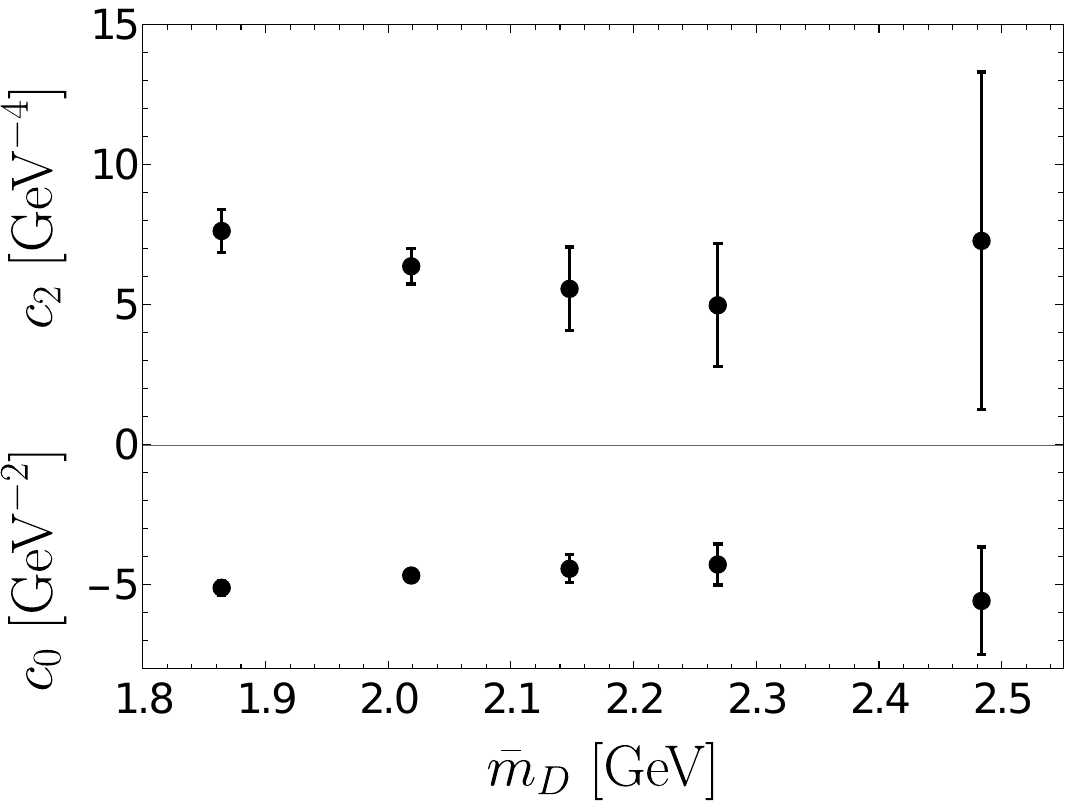}}
\caption{The dependence of the fitted parameters of the short range potential $c_0$ (lower panel) and $c_2$ (upper panel) on the spin-averaged mass $\mbar$ (see Table~\ref{tab:results}).}
\label{fig:cs}
\end{figure}

The Lippmann--Schwinger equation also depends on the values of the pion decay constant $f_\pi$ and the $DD^*\pi$ coupling $g_c$ evaluated at $m_\pi\simeq 280$~MeV as provided in Ref.~\cite{Du:2023hlu}. 
The pion mass dependence of these quantities was taken from the one-loop chiral perturbation theory \cite{Gasser:1983yg} result of Ref.~\cite{Baru:2013rta} and the lattice value of the $D^*D\pi$ coupling from Ref.~\cite{Becirevic:2012pf} was used as input. The regularisation scheme employed consists of imposing a sharp cut-off $\Lambda=0.5$~GeV in the loop 3-momentum. The results of the fits are collected in Table~\ref{tab:results} and visualised in Fig.~\ref{fig:pcotdelta}. These best fit estimates lead to a potential consistent with the form discussed above and sketched in Fig.~\ref{fig:Vtot}.
The fitted values of the counter terms shown in Fig.~\ref{fig:cs} support strong attraction at short distances while the long-range part of the one-pion exchange with $\mu_\pi^2>0$ (see Eqs.~\eqref{mupi} and \eqref{VOPE0}) provides weak repulsion at large distances.
See also Appendix~\ref{app:pot} for further discussion of the form of the potential.

\begin{figure}[t!]
\includegraphics[width=0.9\columnwidth]{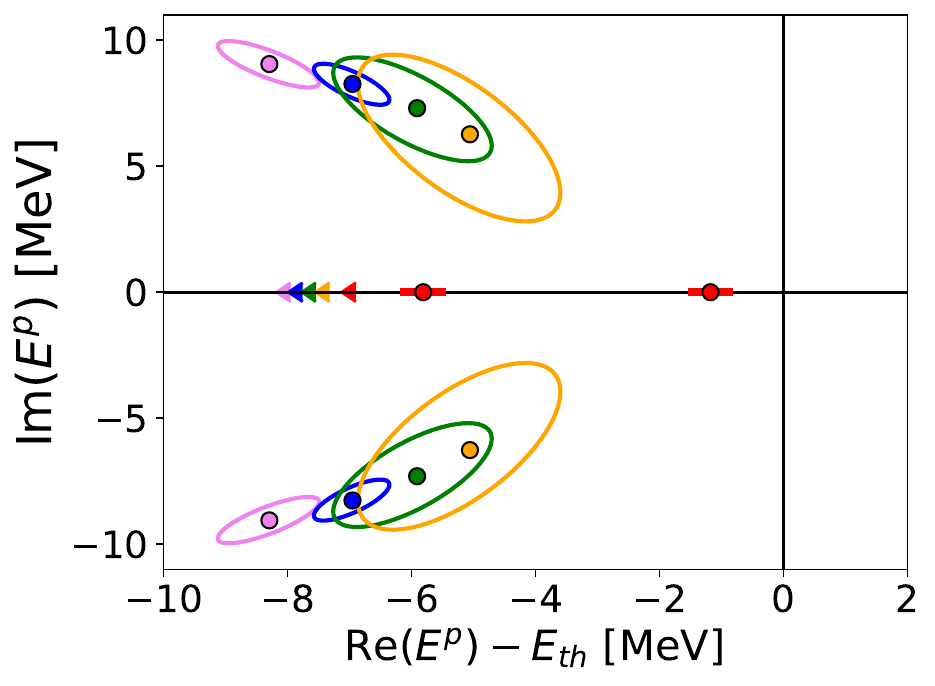}

\vspace{0.3cm}

$\qquad\quad \ $\includegraphics[width=0.9\columnwidth]{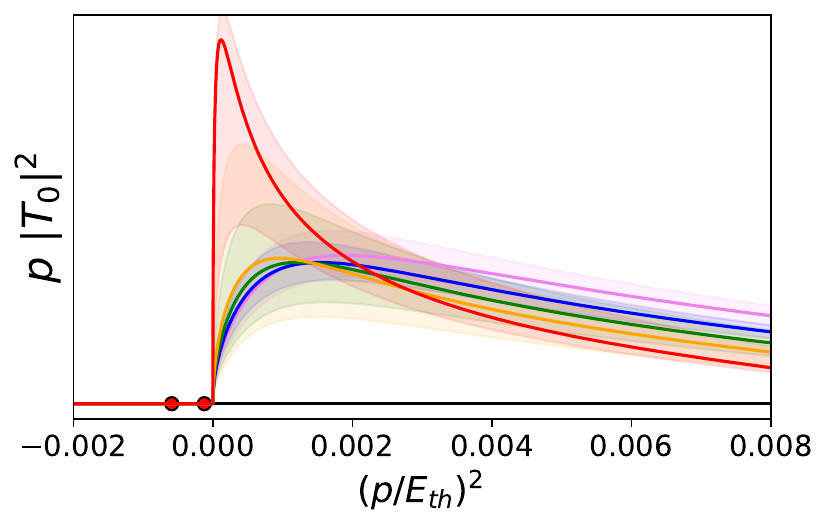}
\caption{The pole trajectories and scattering rate for five heavy-quark masses. Top: The poles of the $DD^*$ scattering amplitude 
(circles with ellipses for 1$\sigma$ uncertainties) and the branch point of the left-hand cut (triangles) quoted in Table~\ref{tab:results}. Bottom: Scattering rate $\propto p |T_0|^2$ in arbitrary units and the location of two virtual poles on the real axes. The heavy-quark mass increases going from the violet to red data points.}
\label{fig:pole}
\end{figure}

\begin{figure*}[t!]
\begin{flushleft}
\hspace*{0.09\textwidth}\includegraphics[width=0.32\textwidth]{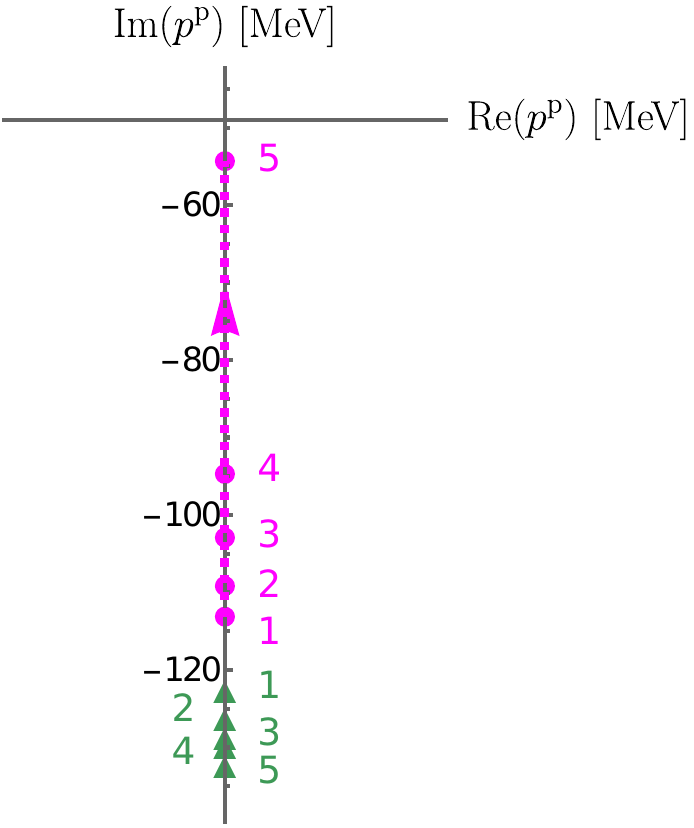}
\hspace*{0.075\textwidth}
\raisebox{20mm}{\includegraphics[width=0.48\textwidth]{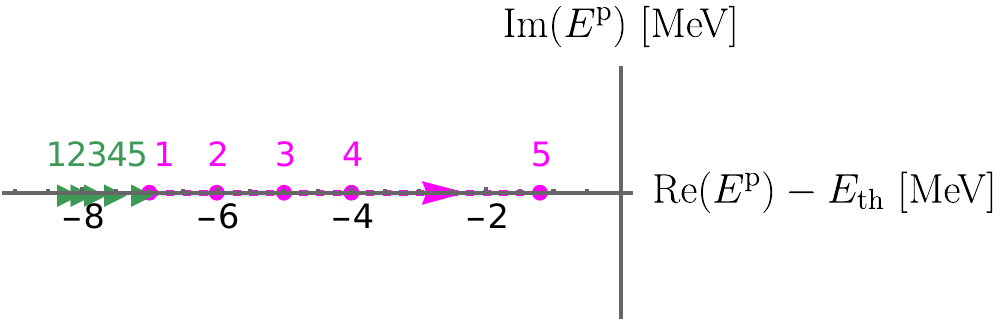}}\\[3mm]
\includegraphics[width=0.49\textwidth]{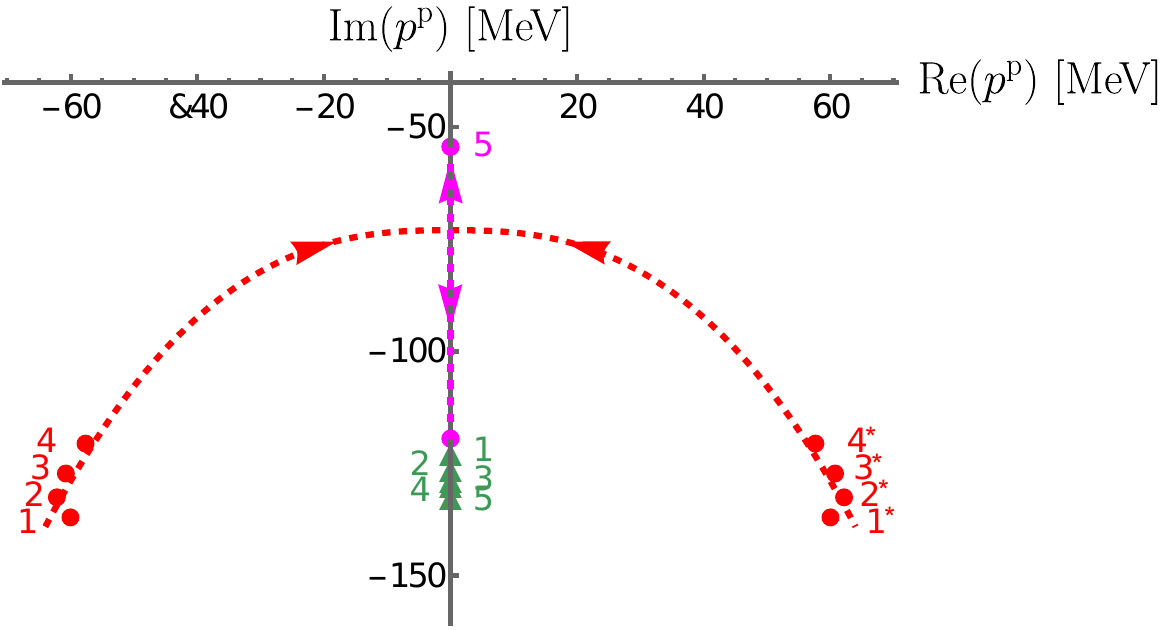}
\raisebox{5mm}
{\includegraphics[width=0.49\textwidth]{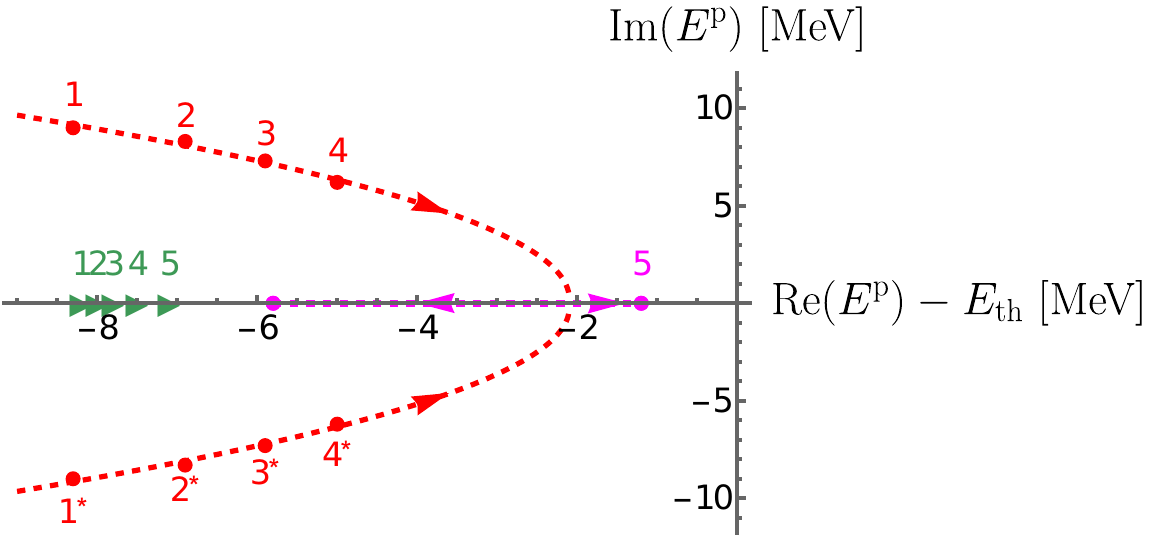}}
\end{flushleft}
\caption{Summary of the poles trajectories (only the central values of the pole positions are shown) in the complex momentum (left) and energy (right) plane for the effective range expansion (upper panel) and effective field theory approach (lower panel). The heavy-quark mass increases from Set 1 to Set 5 as listed in Table~\ref{tab:results}. The resonance and virtual poles are shown as red and magenta points, respectively. The position of the left hand cut branch point is shown by green triangles. The complex conjugated poles are marked with an asterisk. The dashed lines guide the eye for the poles motion between the lattice points (they do not represent a fit or interpolation), with the arrows showing the direction of the motion of the pole as the heavy quark mass increases.}
\label{fig:polespE}
\end{figure*}

Having solved the Lippmann--Schwinger equation in Eq.~(\ref{LS}), for each heavy-quark mass we obtain a scattering amplitude with the fitted values of the counter terms quoted in Table~\ref{tab:results}.\footnote{The analysis in Ref.~\cite{Meng:2023bmz} reaffirms the scattering amplitude for our Set 2 also below the left-hand cut. 
The approach is based on a Hamiltonian formalism employed to extract finite-volume energies in the plane wave basis \cite{Meng:2021uhz} that is free of the issues related to left-hand cuts. Meanwhile, to stay on a safe ground, in the present analysis, we exclude the lattice points below the left-hand cut until they are independently confirmed using the L{\"u}scher formalism adapted to the presence of the left-hand cuts, for example, as outlined in Refs.~\cite{Raposo:2023oru,Raposo:2023nex,Dawid:2023jrj,Hansen:2024ffk}.}
Next we search for the poles of the on-shell scattering amplitude $T(E)$ defined in Eq.~(\ref{amplT}) across the complex energy plane $E$ (see Appendix~\ref{app:LSE} for technical details). 
All poles are found to reside on the second Riemann sheet---we list them in Table~\ref{tab:results}. The pole trajectory with
varying heavy-quark mass is visualised in Figs.~\ref{fig:pole} and \ref{fig:polespE}. In particular,
\begin{itemize}
\item for smaller values of $\mbar$, the physical amplitude possesses a pair of symmetric (complex conjugated) subthreshold resonance poles (see Appendix~\ref{app:pot} for a qualitative discussion).
\item With increasing $\mbar$, these subthreshold resonance poles move towards one another and finally collide on the real axis below the $DD^*$ threshold. Then they turn back to back to each other and move towards the left-hand cut branch point and the $DD^*$ threshold, respectively. 
In this way, the poles trajectories demonstrate a clear pattern of a subthreshold resonance state (two complex conjugated poles) turning to a pair of virtual states (two ``independent'' poles on the real axis below the threshold) with increasing heavy-quark mass. We always conventionally identify the physical state with the pole closest to the $DD^*$ threshold. 
\item The near-threshold poles render an enhancement of the scattering rate $\propto p |T_0|^2$ just above the $DD^*$ threshold, as shown in Fig. \ref{fig:pole}.
\item The observed evolution of the poles with increasing heavy-quark mass demonstrates a clear pattern of a stronger bound system for heavier constituents. This pattern can be qualitatively explained by a weak heavy-quark mass dependence of the total interaction potential. Although, as claimed above, the full potential $V(\vep,\vep')$ from Eq.~\eqref{Vtot} does not have a straightforward representation in the coordinate space, its simplified version studied in Appendix~\ref{app:pot} indeed demonstrates a weak dependence on $m_c$. Under these circumstances, the main dependence of the eigen-energies on the heavy-quark mass stems from the kinetic term of the $DD^*$ pair that scales inversely proportional to $m_c$. We, therefore, conclude that it is this interplay of the kinetic energy (decreasing with increasing $m_c$) and (nearly $m_c$-independent) interaction potential in the system that renders the motion of the poles of the $DD^*$ scattering amplitude discussed above.
\end{itemize}

An important remark is in order here. All the calculations above are performed for the same value of the regulator $\Lambda$ (introduced as a sharp cut-off in the loop 3-momentum). As our aim is to determine the mass-dependence of the contact terms $c_0$ and $c_2$ and extract the pole positions we do not investigate the dependence of the results on $\Lambda$. However, the $\Lambda$-dependence of these unrenormalised contact terms is implicitly understood and a naive use of the numerical values of the $c_0$ and $c_2$ provided in this work in a different regularisation scheme without refitting the lattice data would lead to uncontrolled results.

\subsection{Light-quark mass dependence}

Our present and previous \cite{Padmanath:2022cvl} results correspond to $m_\pi\simeq 280$~MeV 
while the lattice $DD^*$ scattering amplitude was also extracted in Refs.~\cite{Lyu:2023xro,Chen:2022vpo} for pion masses $m_\pi\simeq 146$ and 348~MeV, respectively, at the physical charm quark mass. We re-analysed these lattice data also using the effective field theory approach, following the same strategy as for our data. However, the available lattice data correspond to significantly different energy ranges around the threshold and several relevant cautionary remarks are in order. Thus, since no direct comparison with our results is possible, we present these analyses in Appendix~\ref{app:other}. The general conclusion we deduce from this study is a more attractive interaction between $D$ and $D^*$ for lighter pions and, as a consequence, a stronger bound $DD^*$ system that is expected to reproduce the experimentally observed $\Tcc$ in the physical limit.

\begin{figure}[t]
\centerline{\includegraphics[width=0.5\columnwidth]{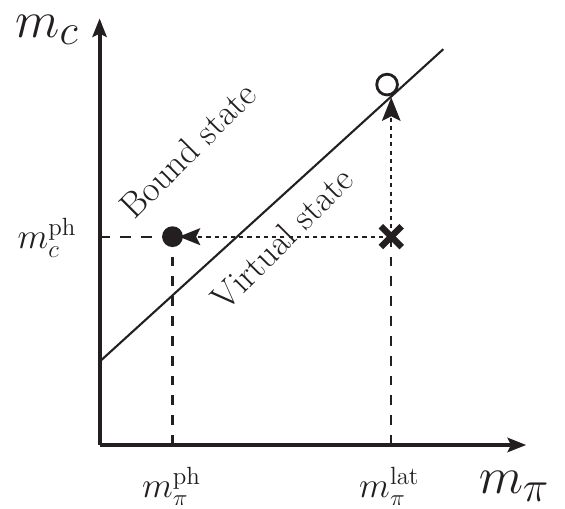}
\includegraphics[width=0.5\columnwidth]{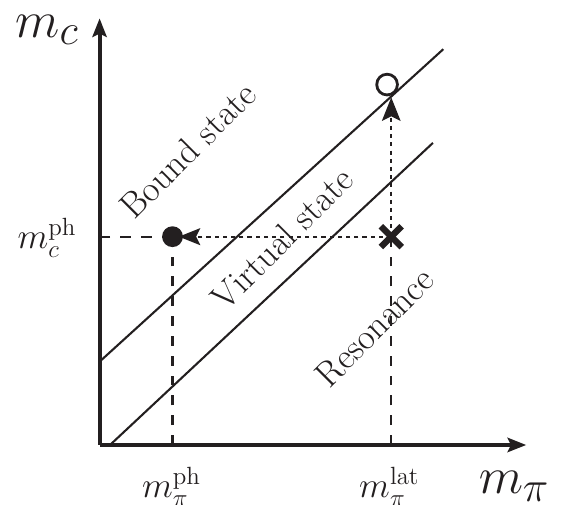}
}
\caption{Sketch of the motion of the $DD^*$ scattering amplitude pole in the $(m_\pi,m_c)$ plane for the effective range expansion (left plot) and effective field theory (right plot) approaches. The filled and open circles indicate, respectively, the position of the bound state pole in the physical limit and that anticipated on the lattice for a sufficiently large heavy-quark mass.}
\label{fig:erevseft}
\end{figure}

\section{Conclusions}
\label{sec:conclusions}

In this work, we studied the subthreshold pole locations in the $DD^*$ scattering amplitude using lattice QCD data with a pion mass $m_\pi\simeq 280$~MeV and several values of the heavy-quark mass, which lie both below and above the physical charm quark mass. In all cases, for kinematical reasons, the $D^*$ is stable with respect to its strong decay to $D\pi$, and the radiative mode $D^*\to D\gamma$ is absent.

We employ two types of analysis. The first procedure utilises an effective range expansion for the energy dependence of the $DD^*$ scattering amplitude in the elastic region above threshold. Ignoring potentially nonnegligible left-hand cut effects and naively extrapolating the resultant energy dependence below the threshold renders a virtual state along the real axis below threshold for all the heavy-quark masses (corresponding to $DD^*$ reduced masses in the range $\muD\simeq 0.9-1.2$~GeV). The observed quark mass dependence of the effective range parameters and the resultant pole are summarised in 
Figs.~\ref{fig:results-ere}-\ref{fig:ere-expectation}. The virtual pole approaches the threshold as the heavy-quark mass is increased and is expected to become a bound state, related to $\Tcc$, at $\muD^{\rm crit}=1.35(^{+17}_{-13})$~GeV. 

The second analysis is based on the effective field theory framework and the $DD^*$ scattering amplitude is obtained from the numerical solution of the Lippmann--Schwinger equation. The interaction potential includes two contact terms parametrising the short-range part of the potential to order ${\cal O}(p^2)$ and the pion exchange term, which is parameter-free and incorporates the effects arising from the left-hand cut. The dependence of the two contact terms fitted to the lattice data is shown in Fig.~\ref{fig:cs}. Their resulting values imply significant attraction at small distances for all heavy-quark masses. At the same time, for our kinematics with $m_\pi>m_D+m_{D^*}$, the pion exchange implies a slight repulsion at larger distances. As the heavy-quark mass increases, the pole representing the ``physical'' state evolves from a subthreshold resonance in the complex energy plane to a virtual state. It is expected to turn into a bound state for heavier constituents. We summarise the pictures of the poles motion obtained in both approaches employed in this work in Fig.~\ref{fig:polespE}.

Our analysis favours weak overall dependence of the $D$-$D^*$ interaction on the heavy-quark mass and stronger bound systems for heavier constituents. This is in agreement with general expectations that the binding energy of a doubly-heavy system mainly scales with the kinetic term that is inversely proportional to the heavy mass.
The estimated value of the critical reduced mass from Eq.~(\ref{critERE}), when the virtual state converts to a bound state, is very close to the reduced mass of the physical $B^*D$ system $\muD^{\rm ph}(B^*D) \simeq 1.38$~GeV relevant for the $\Tbc$ state with $J^P=1^+$. This observation hints towards the existence of the $\Tbc$ pole (bound or virtual) very near the $B^*D$ threshold. However, this insight needs to be taken with caution since our study explicitly considers two heavy quarks of the same flavour and relies on the effective range expansion, which is not justified in the presence of a near-by left-hand cut. A similar estimate based on the effective theory approach is ambiguous. On the one hand, the dependence of $c_0$ and $c_2$ on the spin average $D$ meson mass in Fig.~\ref{fig:cs} is consistent (in the sense of a $\chi^2$ fit) with $1/\mbar$ and $1/\mbar^2$, respectively.\footnote{We note that, for the $X(3872)$, the $1/\mbar$ mass scaling of the leading-order contact term was proposed in Ref.~\cite{AlFiky:2005jd} based on demanding a proper power counting for a heavy-heavy system. However, according to the claim of Ref.~\cite{Baru:2018qkb}, such a dependence (as well as any other) cannot be derived directly from effective field theory.} On the other hand, in both plots, the most right point corresponding to the largest heavy-quark mass has a large uncertainty and is basically ignored by the fit. Under these circumstances, the behaviour of the contact terms cannot be reliably extrapolated to larger values of $\mbar$ consistent with the $\Tbc$ case. Thus, at the present stage we cannot make a definite quantitative conclusion concerning the nature of the $\Tbc$. Direct lattice studies of this state have not reached complete consensus yet as to whether it would be a shallow bound state \cite{Padmanath:2023rdu,Alexandrou:2023cqg} or not \cite{Hudspith:2020tdf,Meinel:2022lzo}, although the more recent and extensive investigations \cite{Padmanath:2023rdu,Alexandrou:2023cqg} suggest that a shallow bound state is more likely. Meanwhile, we notice that our qualitative inference of $\Tbc$ as a possible shallow bound state within the effective range approximation complies well with the findings of Refs.~\cite{Padmanath:2023rdu,Alexandrou:2023cqg}. This intriguing tetraquark is being experimentally searched for by LHCb.

We conclude that, while the results obtained the employing effective range expansion and effective field theory qualitatively demonstrate similar patterns, the latter approach is generally more rigorous and provides deeper insight into the properties of the system under study. 
For completeness, we perform a similar analysis of the lattice data from Refs.~\cite{Chen:2022vpo,Lyu:2023xro} obtained at different pion masses and arrive at analogous conclusions as those based on our lattice data. We leave a more detailed comparison of these results to future studies. 

In Fig.~\ref{fig:erevseft} we provide a sketch that summarises our understanding of the $\Tcc$ pole motion as a function of the pion mass and the heavy-quark mass in both approaches based on effective range expansion and effective field theory employed above. Meanwhile, the results already obtained allow us to conclude that a specific location of the physical $\Tcc$ in the spectrum must stem from a very delicate fine tuning between the masses of the light quarks and charm quark that takes place in nature. The origins and further implications of this fine tuning still remain to be understood.

\begin{acknowledgments}

The authors would like to thank M. Doering, M.-L. Du, A. Filin, C. Hanhart, M. Hansen, B. Kubis, L. Leskovec, B. Mevlja, M. Mikhasenko, and A. Milstein for valuable discussions. A.N. and S.P. are supported by the Slovenian Research Agency (research core Funding No. P1-0035). A.N. also acknowledges support from the CAS President’s International Fellowship Initiative (PIFI) (Grant No. 2024PVA0004). M.P. gratefully acknowledges support from the Department of Science and Technology, India, SERB Start-up Research Grant No. SRG/2023/001235 and Department of Atomic Energy, India. We thank our colleagues in CLS
for the joint effort in the generation of the gauge field ensembles which form a basis for the 
computation. We use the multigrid solver of
Refs.~\cite{Heybrock:2014iga,Heybrock:2015kpy,Richtmann:2016kcq,Georg:2017diz} for the inversion of the Dirac operator. Our code implementing
distillation is written within the framework of the Chroma software package \cite{Edwards:2004sx}. The simulations were performed on the Regensburg Athene2 cluster. We thank the authors of Ref. \cite{Morningstar:2017spu} for making the~{\it TwoHadronsInBox} package public.  We acknowledge the HPC RIVR consortium (www.hpc-rivr.si) and EuroHPC JU (eurohpc-ju.europa.eu) for funding this research by providing computing resources of the HPC system Vega at the Institute of Information Science (www.izum.si). 

\end{acknowledgments}

\appendix

\section{Discussion of the $DD^*$ interaction potential}
\label{app:pot}

In this appendix we qualitatively discuss the interaction in the $DD^*$ system that, as stated in Sec.~\ref{sec:eft}, in general does not reduce to a simple potential in the coordinate space. Meanwhile, a simplified approach to this interaction adopted here allows us to pick up its most general features relevant for understanding the poles motion.

We start from the central part of the one-pion exchange potential in the momentum space given in Eq.~(\ref{VOPE0}) and move to coordinate space,
\be
V_\pi^{\rm cent}(r)=\frac{g_c^2}{4f_\pi^2}\left(-\delta^{(3)}(\ver)+\frac{\mu_\pi^2}{4\pi r}e^{-\mu_\pi r}\right),
\label{VOPEr}
\ee
where the first term in parentheses on the right-hand side contributes to the short-range interaction in the $DD^*$ system while the second term describes the long-range tail. As discussed in Sec.~\ref{sec:eft}, we notice that the sign of the long-range contribution depends on the sign of the effective parameter $\mu_\pi^2$---in the current lattice settings, $\mu_\pi^2>0$, so the long-range pion exchange is repulsive.

Augmenting the pion exchange interaction (\ref{VOPEr}) with the contact term from Eq.~(\ref{VCT}), the total potential reads
\be
V(r)=\left(2c_0-\frac{g_c^2}{4f_\pi^2}\right)\delta^{(3)}(\ver)+\frac{g_c^2\mu_\pi^2}{16\pi f_\pi^2 r}e^{-\mu_\pi r},
\label{Vtotr}
\ee
where, for the sake of simplicity, the ${\cal O}(p^2)$ contact terms were omitted. It is instructive to notice that, in the current lattice settings with a large $\mu_\pi$, the dimensionless coupling parameter $g_c^2\mu_\pi^2/f_\pi^2$, that defines the strength of the pion exchange at large distances \cite{Fleming:2007rp}, several time exceeds its value in the physical world, so the pion exchange interaction is effectively stronger on the lattice.

We would also like to note that, in the settings of this work, the two contributions to the short-range potential in Eq.~(\ref{Vtotr}) appear to be comparable in size, $2|c_0|\sim g_c^2/(4f_\pi^2)\sim 10~\mbox{GeV}^{-2}$. We emphasise, however, that, according to general principles of effective field theories, these two contributions cannot be disentangled in a scheme-independent way \cite{Baru:2015nea}.

\begin{figure}[t!]
\centerline{\includegraphics[width=0.99\columnwidth]{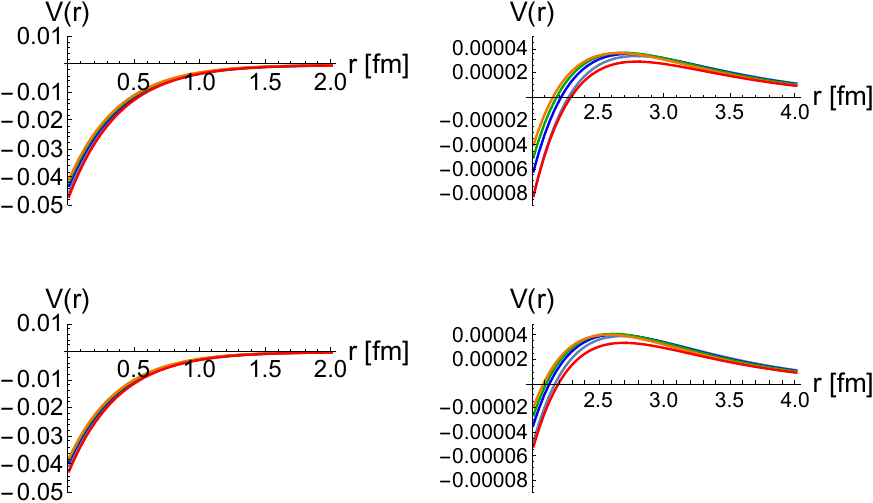}}
\caption{The shape of the potential \eqref{Vreg} between $D$ and $D^*$ as function of their separation $r$ for five different heavy quark masses (the colour coding corresponds to Fig.~\ref{fig:results-ere}). The short-range (left) and long-range (right) parts of the potentials have very different scales, so they are separated to improve the presentation. The plots correspond to $V_{\rm CT}=2c_0$ (top) and $V_{\rm CT}=2c_0 + 4 c_2 (\tfrac{1}{2}\Lambda)^2$ (bottom) as discussed in the text. The plotted curves have the shape sketched in Fig.~\ref{fig:Vtot} and demonstrate a weak dependence on the heavy-quark mass.}
\label{fig:Vr}
\end{figure}

For the discussion of the pole motion in the complex energy plane we also find it useful to perform a study of the interaction potential at realistic values of the parameters and, in particular, observe the potential dependence on the charm quark mass. To this end we study the regularised potential
\be
V_{\rm reg}(r)=\int \frac{d^3q}{(2\pi)^3}e^{i\veq\ver}\Bigl(V_\pi^S(\veq)+V_{\rm CT}\Bigr) f_{\rm reg}(\veq^2),
\label{Vreg}
\ee
with the large-momentum regulator\footnote{Although this regulator, which is more convenient for the semianalytical studies performed in this appendix, differs from the sharp cut-off employed in the numerical studies of the Lippmann--Schwinger equation \eqref{LS}, this difference is irrelevant for the qualitative discussion here.}
\be
f_{\rm reg}(\veq^2)=\left[\frac{\Lambda^2-m_\pi^2}{\Lambda^2-(m_{D^*}-m_D)^2+\veq^2}\right]^2
\label{freg}
\ee
and $\Lambda=0.5$~GeV. In the regularised potential \eqref{Vreg} the delta-function is smeared and the behaviour of the last term, driven by the pion exchange, is tamed in the limit $r\to 0$. A detailed discussion of such a regularisation can be found, for example, in Refs.~\cite{Tornqvist:1993ng,Liu:2008fh,Thomas:2008ja}. The pion exchange potential $V_\pi^S(q)$ corresponds to that from Eq.~\eqref{VOPESS1} with the $D^{(*)}$-meson recoil terms neglected, and the contact interaction $V_{\rm CT}$ from Eq.~\eqref{VCT} is considered for the following two cases: (i) $p=p'=0$ and (ii) $p=p'=\Lambda/2$. The results shown in Fig.~\ref{fig:Vr} demonstrate that all the curves corresponding to the five different charm quark masses very weakly deviate from one another. This weak dependence must provide a natural explanation for the poles motion with the increasing $m_c$, depicted in Figs.~\ref{fig:pole}, that is mainly driven by the kinetic $DD^*$ energy that scales as $1/m_c$. 
Also, all the potentials depicted in Fig.~\ref{fig:Vr} have the shape qualitatively sketched in Fig.~\ref{fig:Vtot}. In particular, we emphasise
a finite (however, regulator-dependent) value of $V_{\rm min}=V(r=0)$ and the existence of a hump at moderate separations needed to smoothly interpolate between the decreasing repulsive long-range one-pion exchange and the attractive interaction at short distances. It is expected, therefore, that such a potential may support not only bound or virtual states but also resonances that correspond to pairs of complex conjugated poles in the complex energy plane---see, for example, Ref. \cite{bzp} for details. The existence of such resonance poles for the $\Tcc$ on the lattice was previously discussed in Ref.~\cite{Du:2023hlu}.

\begin{table*}[t!]
\begin{ruledtabular}
\begin{tabular}{lc|ccc |cc}
Ref.& $m_\pi$ [GeV]& $c_0$~[GeV$^{-2}$] & $c_2$~[GeV$^{-4}$] & $\Ep-\Eth$~[MeV] & $c_0$~[GeV$^{-2}$] & $c_2$~[GeV$^{-4}$] \\
% \hline
 & & \multicolumn{3}{c|}{fit (i)} & \multicolumn{2}{c}{fit (ii)} \\
\hline
\cite{Chen:2022vpo} & 0.348 & $-4.60$& 6.90 & $\simeq~ -10 \pm 12 ~i$ & $-3.86\pm 0.99$ & 0\\
\cite{Chen:2022vpo}$^*$ & 0.348 & $-3.23\pm 0.39$& $6.37\pm 1.48$ & $\simeq~ -15 \pm 19 ~i$ & $-2.00\pm 0.47$ & 0\\
\cite{Padmanath:2022cvl}& 0.280 & $-4.66\pm 0.21$ & $6.38\pm 0.63$& $\simeq~ -6.9\pm 8.3 ~i$ & $-3.21\pm 0.69$ &0 \\
\cite{Lyu:2023xro} & 0.146 & $-11.63\pm 0.14$& $41.15\pm 0.63$& $\in [-1,0]~$ & $-5.89\pm 0.02$ &0 \\
\end{tabular}
\end{ruledtabular}
\caption{The contact terms $c_{0,2}$ and the pole locations extracted using the effective field theory approach from the fits to the lattice data from Refs.~\cite{Chen:2022vpo,Padmanath:2022cvl, Lyu:2023xro}. The corresponding plots are presented in Fig.~\ref{fig:pcotd-mpi} in the same order. Fit (i) has $c_0$ and $c_2$ as free parameters while in fit (ii) only $c_0$ is varied with $c_2=0$. The asterisk indicates the results of the fit to the generated pseudo-data as explained around Eq.~\eqref{pseudo-fit}.
To facilitate the discussion in the text, for fit (i) we also provide the pole position $\Ep-\Eth$. Since, in view of the caveats explained in the text, the uncertainties of the pole positions for the alternative lattice data sets cannot be reliably determined, they are not quoted. The virtual pole based on the data from Ref.~\cite{Lyu:2023xro} remains within $\simeq 1$~MeV below the $DD^*$ threshold. }
\label{tab:pcotd-mpi}
\end{table*}

\section{Approach to solving Lippmann--Schwinger equation (\ref{LS})}
\label{app:LSE}

In this appendix, we provide some details of the approach to solving Lippmann--Schwinger equation (\ref{LS}), along the lines of Ref.~\cite{Du:2023hlu}. Since one-pion exchange does not contain any free parameters (which would need to be determined by fitting to the lattice data), we first solve Lippmann--Schwinger equation for the one-pion exchange potential alone,
\begin{align}
T_\pi(\vp,\vp';E)&=V_\pi(\vp,\vp')\nonumber\\[-2mm]
\label{LS1}\\[-2mm]
&-\int \dk V_\pi(\vp,\vk) G(\vk;E)T_\pi(\vk,\vp';E),\nonumber
\end{align}
to obtain the amplitude $T_\pi(\vp,\vp';E)$. 

To proceed, the contact potential in Eq. (\ref{VCT}) can be conveniently written in a separable matrix form as
\be
V_{\rm CT}(\vp,\vp')=f^T(\vp)\hat{C}f(\vp'),
\ee
where $\hat{C}$ is a constant matrix,
\be
\hat{C}=
\begin{pmatrix}
2c_0 & 2c_2\\
2c_2 & 0
\end{pmatrix},
\ee
and $f(\vp)$ and $f^T(\vp)$ is a momentum-dependent vector and its transpose, respectively,
\be
f(\vp)=
\begin{pmatrix}
1\\p^2
\end{pmatrix}
,\quad f^T(\vp)=(1\ p^2).
\ee
The solution of the full Lippmann--Schwinger equation in Eq. (\ref{LS}) can be then constructed as \cite{Kaplan:1996xu}:
\be
T(\vp,\vp';E)=T_\pi(\vp,\vp';E)+X^T(\vp;E)\hat{t}(E)X(\vp';E),
\label{Tsol}
\ee
where
\begin{align*}
X(\vp;E)&=f(\vp)-\int\dk f(\vk)G(\vk;E)T_\pi(\vk,\vp;E),\\
X^T(\vp;E)&=f^T(\vp)-\int\dk T_\pi(\vp,\vk;E)G(\vk;E)f^T(\vk),
\end{align*}
and
\be
\hat{t}^{-1}(E)=\hat{C}^{-1}+\int\dk f(\vk)G(\vk;E)X^T(\vk;E).
\label{tm1}
\ee
Since the unknown low-energy constants $c_0$ and $c_2$, treated as fitting parameters, enter relation in Eq. (\ref{tm1}) only, all integrals contained in the expressions above can be precalculated, and the fitting procedure becomes particularly simple.

To search for the poles of the amplitude (\ref{Tsol}) in the complex energy plane $E$, we set $p=p'$ such that $E=\Eth +p^2/(2\muD)$ and resort to the standard definition of the Riemann sheets. Namely, we define the momentum on the physical Riemann sheet as
\be
p=\frac{1}{2E}\lambda^{1/2}(E^2,m_D^2,m_{D^*}^2),
\ee
with $\lambda$ for the K{\"o}llen triangle function,
$$
\lambda(x,y,z)=x^2+y^2+z^2-2xy-2xz-2yz.
$$
Then we find (with the subscripts $_I$ and $_{II}$ for the first and second Riemann sheet, respectively) that, across the unitary cut,
\be
G_I(E+i0)-G_I(E-i0)=2i\mbox{Im}G_I(E+i0)=2ip
\label{discG}
\ee
and
\be
G_{II}(E-i0)=G_I(E+i0)=G_I(E-i0)+2ip.
\label{GIII}
\ee
Since the potential does not have a discontinuity on the real axis above the two-body cut branch point then Eq.~(\ref{GIII}) implies the relation
\be
T^{-1}_{II}(E-i0)=T^{-1}_I(E+i0)=T^{-1}_I(E-i0)+i\frac{\muD p}{\pi},
\ee
which we employ in the pole search on the second Riemann sheet of the complex energy plane.

\section{Comment on the light-quark mass dependence}
\label{app:other}

In this appendix, we analyse the lattice data on the $\Tcc$ state at the physical charm quark mass and $m_\pi\simeq 146$ \cite{Lyu:2023xro} and 348~MeV \cite{Chen:2022vpo}. We employ the same effective-field-theory-based approach as that used in Sec.~\ref{sec:results} on our lattice data at $m_\pi\simeq 280$~MeV. For the reasons explained below, we perform two types of fits for the counter terms:
\begin{itemize}
\item fit (i): two fitting parameters, $c_0$ and $c_2$;
\item fit (ii): one fitting parameter, $c_0$, with fixed $c_2=0$.
\end{itemize}

\begin{figure}[t!]
\centerline{\includegraphics[width=0.99\columnwidth]{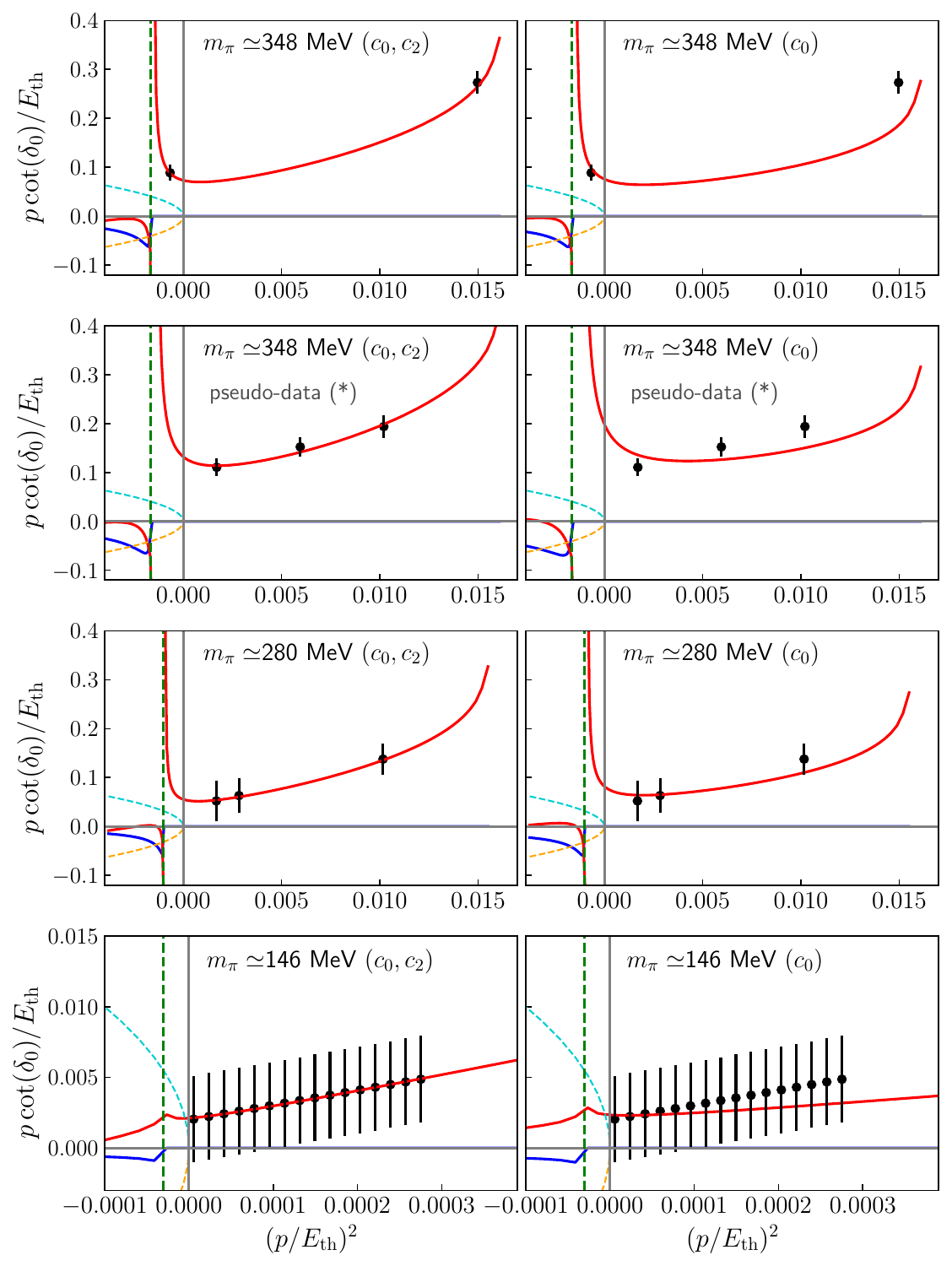}}
\caption{Behaviour of the function $\pcot$ based on the lattice data at $m_\pi \simeq 348$~MeV \cite{Chen:2022vpo}, 280~MeV \cite{Padmanath:2022cvl}, and 146~MeV \cite{Lyu:2023xro}. All the data are analysed using the effective field theory approach described in Sec.~\ref{sec:results}. Left panel: fit (i) with both parameters $c_0$ and $c_2$ fitted to the data. Right panel: fit (ii) with one fitting parameter $c_0$ and $c_2=0$. See Table~\ref{tab:pcotd-mpi} for the fitted values of the counter terms and the poles presented in the same order. The asterisk in the second row indicates the fit to the generated pseudo-data from Ref.~\cite{Chen:2022vpo} as explained around Eq.~(\ref{pseudo-fit}). }
\label{fig:pcotd-mpi}
\end{figure}

\subsection{Our lattice data at $m_\pi\simeq 280$~MeV}

For the sake of comparison, in Table~\ref{tab:pcotd-mpi} and Fig.~\ref{fig:pcotd-mpi}, we provide the fits to our lattice data for Set 2 that corresponds to the heavy-quark mass closest to the physical charm quark mass. For fit (i) the results are identical to that presented in Table~\ref{tab:results} and shown in Fig.~\ref{fig:pcotdelta}.

\subsection{Lattice data at $m_\pi\simeq 348$~MeV}
 
The data from the CLQCD collaboration~\cite{Chen:2022vpo} correspond to
\begin{align}
&m_\pi=0.348~\mbox{GeV},\quad\mbar =1.986~\mbox{GeV},\nonumber\\[-2mm]
\label{mbarcl}\\[-2mm]
&m_D=1.882~\mbox{GeV},\quad m_{D^*}=2.021~\mbox{GeV}.\nonumber
\end{align}
Only two data points of the four presented in Ref.~\cite{Chen:2022vpo} meet the criteria to lie above the left-hand cut and correspond to the momenta $p\lesssim\Lambda=0.5$~GeV. Thus, strictly speaking, only these two points can be used in the present analysis. This motivates us to employ two different fitting strategies:
\begin{itemize}
\item fit the two data points directly: the results are shown in the first row of Fig.~\ref{fig:pcotd-mpi} and Table~\ref{tab:pcotd-mpi}.
\item Generate and then fit pseudo-data: we use the values of the $I=0$ scattering length and effective range from Ref.~\cite{Chen:2022vpo},
\be
a_0=0.538(33)~\mbox{fm},\quad r_0=0.99(11)~\mbox{fm},
\label{pseudo-fit}
\ee
to generate three pseudo-lattice data points in the momentum range consistent with Set 2 of our lattice data. The results are given in the second row of Fig.~\ref{fig:pcotd-mpi} and Table~\ref{tab:pcotd-mpi} (marked with an asterisk).
\end{itemize}

The amplitudes at both $m_\pi\simeq 348$ and 280~MeV possess a pair of complex conjugated poles listed in Table~\ref{tab:pcotd-mpi} that represent a resonance in the complex energy plane.
It is in line with the discussion of the pole motion presented in Sec.~\ref{sec:results}. We refrain from a more detailed comparison of our results with those from Ref.~\cite{Chen:2022vpo} because the energy ranges covered by the data are different.

\subsection{Lattice data at $m_\pi\simeq 146$~MeV}

\begin{figure}[t!]
\includegraphics[width=0.42\textwidth]{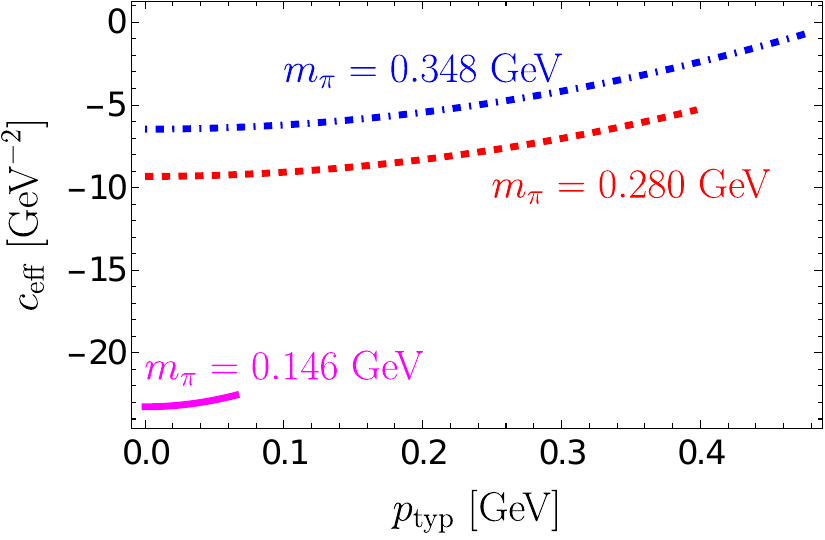}
\caption{The behaviour of the effective contact term defined in Eq.~(\ref{ceff}) for the 3 different pion masses: the blue doted-dashed line is for $m_\pi=0.348$~GeV (the second row in Table~\ref{tab:pcotd-mpi}), the red dashed line is for $m_\pi=0.280$~GeV (the third row in Table~\ref{tab:pcotd-mpi}), and the magenta solid line is for $m_\pi=0.146$~GeV (the last row in Table~\ref{tab:pcotd-mpi}). For each curve, the range for $p_{\rm typ}$ is limited from above by the point from the corresponding lattice data set with the largest momentum $p\lesssim\Lambda=0.5$~GeV.}
\label{fig:ceff}
\end{figure}

We now turn to the data from the HALQCD collaboration~\cite{Lyu:2023xro} that correspond to
\begin{align}
&m_\pi=0.146~\mbox{GeV},\quad\mbar =1.983~\mbox{GeV}, \nonumber \\[-2mm]
\\[-2mm]
&m_D=1.878~\mbox{GeV},\quad m_{D^*}=2.018~\mbox{GeV}.\nonumber
\end{align}
The HALQCD technique is employed in this work, so a disclaimer is in order concerning our re-analysis of the lattice data. Unlike the direct determination of $\pcot$ via the L{\"u}scher's method, the HALQCD approach relies on extraction of the $DD^*$ interaction potential from lattice data. The latter is approximated by a suitable analytic form and then a Schr{\"o}dinger equation is solved. Such an approach involves a certain built-in regularisation of the potential at short distance, which may not be consistent with the sharp cut-off regularisation of the loop integrals (with $\Lambda=0.5$~GeV) employed in this work.
Also, the interaction potential in Eq.~(\ref{Vtot}) derived in the effective field theory framework does not have a straightforward representation as a single-argument function in coordinate space, so reducing the Lippmann--Schwinger equation in Eq.~(\ref{LS}) with such a potential to a Schr{\"o}dinger equation in coordinate space requires additional approximations. 
Finally, the data for $\pcot$ provided in Ref.~\cite{Lyu:2023xro} correspond to a very limited energy range near the $DD^*$ threshold compared with that spanned by our data (see the magenta solid curve in Fig.~\ref{fig:ceff}).

Nevertheless, for completeness, we perform fits to the HALQCD data for $\pcot$ based on the effective field theory technique from Sec.~\ref{sec:eft}. As given above, we either fit both $c_0$ and $c_2$ or fix $c_2=0$ and fit only $c_0$. The results are presented in the last rows of Fig.~\ref{fig:pcotd-mpi} and Table~\ref{tab:pcotd-mpi}. One can conclude that both fits provide a similarly good description of the data within the uncertainties and predict $\Tcc$ to be a very shallow virtual state, in agreement with the claim of Ref.~\cite{Lyu:2023xro}.

\begin{figure}[t!]
\centerline{\includegraphics[width=0.99\columnwidth]{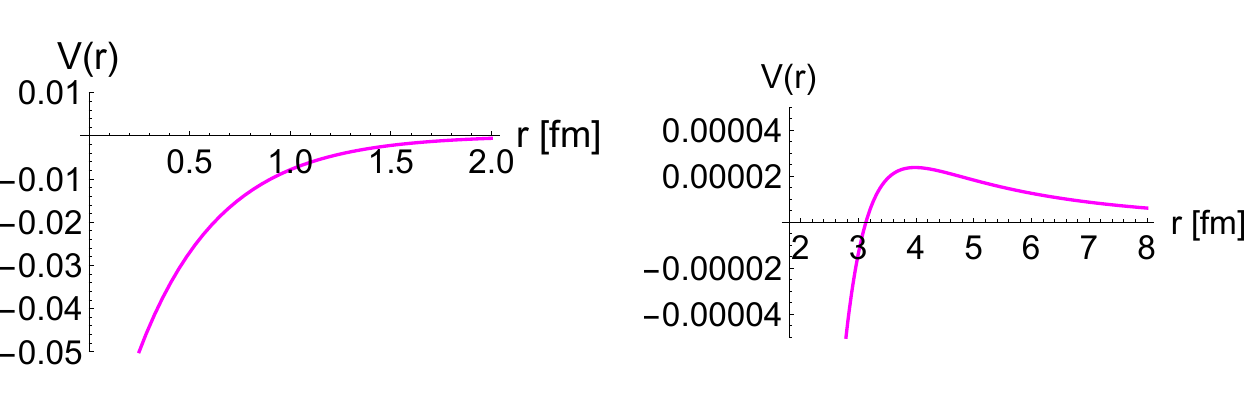}}
\caption{The shape of the regularised potential from Eq.~\eqref{Vreg} between $D$ and $D^*$ based on our fit of the HALQCD lattice data at $m_\pi \simeq 146$~MeV. This fit for $\pcot$ employed one free parameter $c_0$ (see fit (ii) in Table \ref{tab:pcotd-mpi}). The one-pion exchange contribution $\propto e^{-\mu_\pi r}/r$ at large distances (with $\mu_\pi\simeq 43~$MeV for the HALQCD dataset)
is responsible for a slight repulsion.}
\label{fig:Vr-halqcd}
\end{figure}

The shape of the potential $V(r)$ from Eq.~\eqref{Vreg} between $D$ and $D^*$ based on our fit to the HALQCD lattice data is shown in Fig.~\ref{fig:Vr-halqcd}. This potential is attractive for $r\lesssim 2$~fm as observed also by HALQCD. At larger distances our effective field theory approach predicts a slight repulsion that seems to be possibly visible also in the HALQCD potential. This may present a signature of a one-pion exchange contribution with $V(r) \propto e^{-\mu_\pi r}/r$ at large distances. 

\subsection{Comparing the results with different pion masses}

The short-range potential parametrised by the single counter-term $c_0$ stays negative and, as a general pattern, increases as the pion mass decreases, as quoted in Table~\ref{tab:pcotd-mpi} for fit (ii). This supports a stronger attraction in the $DD^*$ system with a decreasing pion mass.

To facilitate the comparison for the two-parameter fits (i) with both $c_0$ and $c_2$ present, we define an effective contact term ({\em c.f.} Eq.~(\ref{VCT}) with $p=p'=p_{\rm typ}$) as
\be
c_{\rm eff}(p_{\rm typ})=2c_0+4c_2 p_{\rm typ}^2
\label{ceff}
\ee
and plot this function in Fig.~\ref{fig:ceff} for the three sets of parameters $\{c_0,c_2\}$ quoted in Table~\ref{tab:pcotd-mpi}.
One can see from this figure that the pattern of a stronger attraction in the $DD^*$ system for smaller $m_\pi$'s observed in fit (ii) above persists in fit (i), too.

Thus, the binding in $\Tcc$ is expected to be stronger for lighter pions. It is consistent with the nature of the physical $\Tcc$, which is found to be a shallow bound state (see, for example, Refs.~\cite{LHCb:2021auc,Albaladejo:2021vln,Du:2021zzh}).

However, we refrain from a direct comparison of our results obtained in this work with those, for example, from Ref.~\cite{Du:2021zzh} where only the contact term $c_0$ was fitted to the experimental data for the physical $\Tcc$ line shape in the channel $D^0D^0\pi^+$ provided in Ref.~\cite{LHCb:2020bwg}. Indeed, although the lattice data studied in this work demonstrate the anticipated pattern of the $\Tcc$ pole converging to a shallow bound state with the pion mass approaching its physical value, in this limit, the $DD^*$ system appears to be sensitive to delicate details of the kinematics and the structure of the cuts. 

\bibliographystyle{apsrev4-1}
\bibliography{Tcclat.bib}

\end{document}